\documentclass[showpacs, oneside, twocolumn, prd, amsmath, amssymb, nofootinbib, superscriptaddress]{revtex4-1}

\usepackage{cases}
\usepackage{amsmath}
\usepackage{amssymb}
\usepackage{amsfonts}
\usepackage{amssymb}
\usepackage{dcolumn}
\usepackage{bm}
\usepackage{bbm}
\usepackage{graphicx}
\usepackage{xcolor}
\usepackage{array}
\usepackage{subfigure}
\usepackage{hyperref}
\usepackage{wasysym}

\newcommand{\be}{\begin{equation}}
\newcommand{\ee}{\end{equation}}
\newcommand{\ba}{\begin{eqnarray}}
\newcommand{\ea}{\end{eqnarray}}
\newcommand{\nn}{\nonumber}
\def\l{\left}
\def\r{\right}

\newcommand{\gsim}{\mathrel{\hbox{\rlap{\lower.55ex \hbox {$\sim$}}
			\kern-.3em \raise.4ex \hbox{$>$}}}}
\newcommand{\lsim}{\mathrel{\hbox{\rlap{\lower.55ex \hbox {$\sim$}}
			\kern-.3em \raise.4ex \hbox{$<$}}}}

\def\bl#1\el{\begin{align}#1\end{align}}

\hypersetup{colorlinks=true,
	breaklinks=true,
	pdfstartview=Fit,
	linkcolor=blue,
	citecolor=blue,
	urlcolor=blue}

\begin{document}

\title{Dirac-Born-Infeld realization of sound speed resonance mechanism for primordial black holes}

\author{Chao Chen}
\email{cchao012@mail.ustc.edu.cn}
\affiliation{Department of Astronomy, School of Physical Sciences, University of Science and Technology of China, Hefei, Anhui 230026, China}
\affiliation{CAS Key Laboratory for Researches in Galaxies and Cosmology, University of Science and Technology of China, Hefei, Anhui 230026, China}
\affiliation{School of Astronomy and Space Science, University of Science and Technology of China, Hefei, Anhui 230026, China}

\author{Xiao-Han Ma}
\email{mxh171554@mail.ustc.edu.cn}
\affiliation{Department of Astronomy, School of Physical Sciences, University of Science and Technology of China, Hefei, Anhui 230026, China}
\affiliation{CAS Key Laboratory for Researches in Galaxies and Cosmology, University of Science and Technology of China, Hefei, Anhui 230026, China}
\affiliation{School of Astronomy and Space Science, University of Science and Technology of China, Hefei, Anhui 230026, China}

\author{Yi-Fu Cai}
\email{Corresponding author: yifucai@ustc.edu.cn}
\affiliation{Department of Astronomy, School of Physical Sciences, University of Science and Technology of China, Hefei, Anhui 230026, China}
\affiliation{CAS Key Laboratory for Researches in Galaxies and Cosmology, University of Science and Technology of China, Hefei, Anhui 230026, China}
\affiliation{School of Astronomy and Space Science, University of Science and Technology of China, Hefei, Anhui 230026, China}

\begin{abstract}
We present a concrete realization of the sound speed resonance (SSR) mechanism for primordial black hole (PBH) formation within a specific model of Dirac-Born-Infeld (DBI) inflation. We perform a perturbative approach to phenomenologically construct such a viable DBI inflation model that involves the nonoscillating stage and the oscillating stage, with a type of specific forms of the warp factor and the potential. We show that the continuous but nonsmooth conjunction of sound speed between two stages does not yield manifest effects on the phenomenology of SSR, and thus, our model gives rise to the same PBH mass spectrum as the original predictions of SSR. Additionally, we also demonstrate that the violation of adiabaticity of the Mukhanov-Sasaki equation does not affect the comoving curvature perturbation after Hubble crossing in the nonresonant region. Making use of observational data, we derive various cosmological constraints on the parameter space. Our analyses show that the predicted tensor-to-scalar ratio is typically small, while the amplitude of primordial non-Gaussianity can meet with cosmic microwave background bounds, and additionally, the consistency relation for single-field slow-roll inflation is softly violated in our case due to the small sound speed variations. 
\end{abstract}


\maketitle

\section{Introduction}
\label{sec:intro}

Primordial black holes (PBHs) may be formed from density fluctuations in the very early Universe \cite{Zeldovich:1966, Hawking:1971ei, Carr:1974nx}, which can be tested through their effects on a variety of cosmological and astronomical processes. In this regard, PBHs can serve as an inspiring tool to probe physics in the very early Universe \cite{Khlopov:2008qy, Sasaki:2018dmp}. In particular, PBHs could be a candidate for (a fraction of) dark matter (DM), which has drawn a lot of attention \cite{Carr:2016drx, Carr:2018poi}. With various forthcoming experimental facilities in gravitational-wave (GW) astronomy, the GW survey has become a promising window to reveal physical processes of PBH formation. There are already many works upon GWs associated with PBHs, for instance, GWs generated by PBH mergers \cite{Sasaki:2016jop, Mandic:2016lcn, Wang:2016ana}, and the induced GWs from the enhanced primordial density perturbations associated with PBH formation \cite{Baumann:2007zm, Ananda:2006af, Kohri:2018awv, Bartolo:2018rku, Unal:2018yaa, Cai:2018dig, Cai:2019jah, Fu:2019vqc, Inomata:2019yww, DeLuca:2019ufz}. Some high-density regions of the very early Universe are expected for PBH formation. One possibility is that there were large primordial inhomogeneities and the resulting overdense regions might collapse to form PBHs \cite{Carr:1975qj}. This motivates many studies of generating PBHs, which require a power spectrum of primordial density perturbations to be suitably large on certain scales that are associated with a particularly tuned background dynamics of quantum fields in the very early Universe (e.g. see \cite{GarciaBellido:1996qt, Garcia-Bellido:2017mdw, Domcke:2017fix, Kannike:2017bxn, Carr:2017edp, Pi:2017gih, Ballesteros:2017fsr, Hertzberg:2017dkh, Motohashi:2017kbs, Franciolini:2018vbk, Biagetti:2018pjj, Ballesteros:2018wlw, Germani:2018jgr, Kamenshchik:2018sig, Georg:2019jld, Dalianis:2019asr, Fu:2019ttf} for studies within inflation, see \cite{Carr:2011hv, Chen:2016kjx, Quintin:2016qro, Clifton:2017hvg} for discussions within bounce, and see \cite{Sasaki:2018dmp} for recent comprehensive reviews).

Recently, a novel mechanism for PBH formation by virtue of sound speed resonance (SSR) was proposed in \cite{Cai:2018tuh}, where it was found that an oscillating sound speed squared can yield nonperturbative parametric amplification on certain perturbation modes during inflation. Accordingly, the power spectrum of primordial density perturbations can have a narrow major peak on small scales, while it remains nearly scale-invariant on large scales as predicted by inflationary cosmology. Several minor peaks of the power spectrum on smaller scales are also predicted in this mechanism and can yield secondary contributions. As a result, the formation of PBHs caused by the resulting peaks in SSR can be much efficient. Moreover, it was found in \cite{Cai:2019jah} that the GWs induced within SSR at the sub-Hubble scales during inflation could become crucial at critical frequency band due to a narrow resonance effect, and hence the spectrum of GWs with double peaks is typically predicted. Additionally, the SSR mechanism can also be generalized to the inflaton-curvaton mixed scenario \cite{Chen:2019zza}, in which the curvaton propagates with a time-oscillating sound speed during inflation, while the inflaton leads to the standard adiabatic perturbations. 

So far, the underlying physics as well as the model realization of SSR, however, are not yet clear. Therefore, in this paper we perform a preliminary investigation on the phenomenological realization of SSR. The nontrivial sound speed is a distinctive feature of noncanonical inflationary scenarios, e.g. k-inflation \cite{ArmendarizPicon:1999rj, Garriga:1999vw}, Dirac-Born-Infeld (DBI) inflation \cite{Silverstein:2003hf, Alishahiha:2004eh} and so on. Specifically, we consider a DBI type of inflation models, which are inspired by string theory with the inflaton field being regarded as the radial position of branes moving inside a warped throat. The scenario of DBI inflation requires the velocity of inflaton to be restricted by the combined effects of the speed limit inherent in the DBI model and the shape of the inflaton's potential, such as the UV \cite{Silverstein:2003hf, Alishahiha:2004eh} and IR models \cite{Chen:2005ad}. Thus, by allowing the inflaton's sound speed to oscillate for a while during inflation, the specific forms of the warp factor and the potential are expected. Accordingly, the crucial step of our attempt on realizing the SSR mechanism is to pin down these two quantities. To do so, we develop a perturbative approach to search for a viable DBI model in the context of a modified anti-de Sitter (AdS) throat, and the corresponding potential is derived by using the Hamilton-Jacobi formalism. To confront with current observations of primordial power spectra and non-Gaussianities, we can obtain the constraints on the parameter space of this model.

This article is organized as follows. In Sec. \ref{sec:DBI}, we describe semianalytically the background evolution of DBI inflation. Then, we in Sec. \ref{sec:realization} derive the requirements for the warp factor and the evolution of inflaton that allows an oscillating sound speed. A perturbative approach is developed to accomplish the requirements on background dynamics, and the potential is acquired by resorting the Hamilton-Jacobi formalism. Afterwards, we in Sec. \ref{sec:viability} discuss the theoretical viability of the reconstructed DBI realization of the SSR mechanism. After that, we in Sec. \ref{sec:constriant} derive observational constraints of our model by analyzing the power spectra, spectrum index, tensor-scalar ratio and non-Gaussianities. We summarize our results with a discussion in Sec. \ref{sec:conclusion}. 
Throughout the article, we work in natural units $c = \hbar =1$ and the reduced Planck mass is defined as $M_p \equiv 1/\sqrt{8 \pi G}$. Additionally, a dot denotes the cosmic time derivative, a prime denotes the derivative with respect to the inflaton $\phi$, and the notation of a comma means the derivative.

\section{DBI inflation} \label{sec:DBI}

Inflation is a prevailing theoretical paradigm of the very early Universe and is strongly favored by cosmological observations, such as cosmic microwave background (CMB) surveys \cite{Ade:2015lrj, Akrami:2018odb}. However, the microscopic nature of the inflaton remains mysterious. In the standard model of single-field slow-roll inflation, the slow-roll condition requires a sufficiently flat potential to drive the inflationary expansion. Therefore, it is a key question to find a dynamical realization of such a flat potential in fundamental theory. An attractive attempt is to embed inflation into string theory and the corresponding models are roughly separated into two categories, depending on whether inflation is a closed string mode (e.g. K\"{a}hler moduli inflation \cite{Conlon:2005jm}) or an open string mode (e.g. brane inflation \cite{Dvali:1998pa, Kachru:2003sx}, DBI inflation \cite{Silverstein:2003hf, Alishahiha:2004eh}). In particular, the DBI model that yields a deviation of primordial sound speed from unity has attracted numerous phenomenological interest, namely, the applications to the curvaton \cite{Li:2008fma, Zhang:2009gw, Cai:2010rt}, the multiple sound speed propagations \cite{Cai:2008if, Cai:2009hw, Cai:2010wt}, and the interpretation of the hemispherical asymmetry anomaly \cite{Cai:2013gma, Cai:2015xba, Li:2019bsg}.

The DBI action is written as
\be \label{Action_DBI}
 S = \int d^4x \sqrt{-g} \Big[ f(\phi)^{-1} \big( 1 - \sqrt{1 + 2 f(\phi) X} \big) - V(\phi) \Big] ~,
\ee
where $X = - \frac12 g^{\mu\nu} \nabla_\mu\phi \nabla_\nu\phi$, and $f(\phi)$ is the redefined warp factor. For the well-studied AdS throat \cite{Klebanov:2000hb}, $f(\phi) = \lambda / \phi^4$ with $\lambda$ being a positive constant, and the form can be phenomenologically deformed depending on the desired model construction. In the spatially flat Friedmann-Lema\^{i}tre-Robertson-Walker background, there is $X = \dot{\phi}^2/2$ for a homogeneous scalar field $\phi$. Moreover, the homogeneous part of the equation of motion (EoM) for a DBI scalar field can be derived from the action \eqref{Action_DBI} by variational principle: 
\be \label{EOM_DBI}
 \ddot{\phi} + 3 H c_s^2 \dot{\phi} + c_s^3 V'(\phi) + \frac{f'(\phi)}{2 f(\phi)} \l( 1 - \frac{2 c_s^2}{1 + c_s} \r) \dot{\phi}^2 = 0 ~,
\ee 
where $H$ is the Hubble parameter and the sound speed squared is defined as
\be \label{cs}
 c_s^2 = 1 - f(\phi) \dot{\phi}^2 ~,
\ee
measures the propagation speed of the field fluctuations \cite{Garriga:1999vw}. The energy density and the pressure of the DBI field are given by
\begin{align}
 \rho = \frac{\gamma^2}{1 + \gamma} \dot{\phi}^2 + V(\phi) ~,~ P = \frac{\gamma}{1 + \gamma} \dot{\phi}^2 - V(\phi) ~,
\end{align}
where we have introduced the Lorentz factor $\gamma$ as follows:
\be \label{Lorentz_Factor}
 \gamma \equiv \frac{1}{\sqrt{1 - f(\phi) \dot{\phi}^2}} = \frac{1}{c_s} ~,
\ee
which tracks the motion of the mobile brane in a warped throat \cite{Silverstein:2003hf}. Since the proper velocity of the brane is $v_p = \sqrt{f(\phi)} \dot{\phi}$, a large value with $\gamma \gg 1$ corresponds to the relativistic motion of the brane. Oppositely, in the nonrelativistic limit with $f \dot{\phi}^2 \ll 1$, the DBI action \eqref{Action_DBI} reduces back to the standard canonical form, which is the regular single-field slow-roll model with the action being $S = \int d^4x \sqrt{-g} ( X - V(\phi) )$.

The most intriguing feature of the DBI field is that the positivity of the square roots in DBI action \eqref{Action_DBI} and the Lorentz factor \eqref{Lorentz_Factor} impose a constraint upon the time-varying $\phi$:
\be \label{Speed_Limit}
 \dot{\phi}^2 \leq \frac{1}{f(\phi)} ~.
\ee
This constraint is irrespective of the shape of inflaton's potential $V(\phi)$ and only subject to the structure of the warp factor $f(\phi)$. It is easy to see from \eqref{Speed_Limit} that the larger $f(\phi)$ leads to the smaller rolling velocity for $\phi$. For instance, for an AdS-like warp factor, $f(\phi) = \lambda / \phi^4$ becomes large in the IR regime of the throat and hence, inflation could happen near the tip of the throat even with a steep potential of $V(\phi)$ \cite{Chen:2005ad}. The nontrivial sound speed squared $c_s^2$ intrinsically appears in DBI inflation \eqref{cs}, which is expected to yield the SSR phenomenology as shall be discussed.

The Friedmann equations read
\begin{align} 
\label{Friedmann1}
 H^2 &= \frac{1}{3 M_p^2} \Big[ \frac{\gamma^2}{1 + \gamma} \dot{\phi}^2 + V(\phi) \Big]~, \\
\label{Friedmann2}
 \dot{H} &= - \frac{1}{2 M_p^2} \gamma \dot{\phi}^2 ~.
\end{align}
In order to solve the coupled Friedman equations \eqref{Friedmann1} and \eqref{Friedmann2} more conveniently, we resort to the Hamilton-Jacobi formalism \cite{Silverstein:2003hf}, in which the field $\phi$ is regarded as the time variable, and this requires that $\phi$ is monotonic. From now on, all the undetermined functions $(H, V, \gamma, f)$ in the above equations are functions of $\phi$. Note that the EoM \eqref{EOM_DBI} can also be obtained from the above Friedmann equations \eqref{Friedmann1} and \eqref{Friedmann2}, so that one can avoid using the complicated form of the EoM \eqref{EOM_DBI} explicitly in the Hamilton-Jacobi formalism.

Using the relationship $H_{,\phi} \dot{\phi} = \dot{H}$, Eq. \eqref{Friedmann2} becomes
\be \label{H_phi}
H'(\phi) = - \gamma(\phi) \frac{\dot{\phi}}{2 M_p^2} ~.
\ee
In the standard inflationary scenario where $\gamma = 1$, one acquires the relation $H'(\phi) = - \dot{\phi} / 2 M_p^2$. Using Eq. \eqref{Friedmann1}, the potential $V(\phi)$ is given by
\be \label{V_phi}
V(\phi) = 3 M_p^2 H(\phi)^2 - \frac{1}{f(\phi)} \sqrt{1 + 4 M_p^4 f(\phi) [H'(\phi)]^2} + \frac{1}{f(\phi)} ~.
\ee
In the following section, one can see that the phenomenological oscillating sound speed determines the evolution of inflaton $\phi$ by the relationship \eqref{cs} when the specific form of the warp factor $f(\phi)$ is given. Then, the parametrized Hubble parameter $H(\phi)$ is derived from \eqref{H_phi}. Finally, we can obtain the inflaton's potential $V(\phi)$ by plugging the functions $f(\phi)$ and $H(\phi)$ into \eqref{V_phi}. Note that, the inflaton's potential in principle comes from brane tensions and interactions \cite{Baumann:2006th, Bean:2007eh}, but the form of potential is not known in general. In this regard, we think of the inflaton's potential as an undetermined function in our model.

\section{DBI Realization of SSR} \label{sec:realization}

In this section, we expect to construct a viable DBI action that can realize the SSR mechanism by choosing the specific forms of the warp factor $f(\phi)$ and the inflaton's potential $V(\phi)$. 
It is suggested in SSR \cite{Cai:2018tuh, Chen:2019zza} that, the sound speed squared for the inflaton field is time evolving during inflation and is parametrized as follows:
\be \label{ssr}
 c_s^2 = 1 -2 \xi [ 1-\cos(2k_*\tau) ], ~~~~~\text{with}~ \tau>\tau_s ~,
\ee
where $\xi$ is a small dimensionless quantity that measures the oscillation amplitude and $k_*$ is the oscillation frequency. Note that, $\xi < 1/4$ is required such that $c_s^2$ is positively definite, and the oscillation begins at $\tau_s$, where $k_*$ needs to be deep inside the Hubble radius with $|k_*\tau_s|\gg1$. 

To realize the oscillating pattern \eqref{ssr} with DBI inflation, the following matching condition derived from \eqref{cs} and \eqref{ssr} ought to be satisfied:
\bl \label{Matching_1}
 f(\phi) \Big( \frac{d \phi}{d \tau} \Big)^2 =& 2 \xi a(\tau)^2 \big[ 1-\cos(2 k_*\tau) \big] \nn \\ 
 \simeq& \frac{2 \xi}{ (\epsilon - 1)^2} \frac{1-\cos(2 k_*\tau)}{H^2 \tau^2} ~,
\el
where we have adopted the quasi-de Sitter approximation for the background evolution, i.e., the slow-roll parameter is assumed to be a small constant, while the Hubble parameter varies slowly, and then the scale factor behaves as $a(\tau) \simeq 1/ (\epsilon - 1) H \tau$. We show the validity of this approximation in the later discussions. The analytic solutions of $f(\phi)$ and $\phi(\tau)$ can in principle be obtained by solving Eqs. \eqref{EOM_DBI}, \eqref{Friedmann1}, \eqref{H_phi}, and \eqref{Matching_1} simultaneously. However, it is not easy to solve these strongly coupled equations analytically. Also, since the warped geometry is determined by the unknown compactification, the form of the warp factor is not fixed in general, and it is convenient to start with the well-studied AdS warp factor. Therefore, we perform a perturbative approach to solve the inflaton evolution $\phi(\tau)$ approximately, and the AdS throat is allowed to be deformed slightly to yield the oscillating pattern for the sound speed squared. As we shall see below, the numerical results and the semianalytic ones match reasonably well.

In the first step, the constant sound speed squared, $c_s^2 = 1 - 2 \xi$, can be realized in the AdS-like throat when a specific evolution of $\phi(\tau)$ is satisfied. Similar to the matching condition \eqref{Matching_1}, we derive the relation
\be \label{Matching_2}
 \frac{\lambda}{\phi^4} \Big( \frac{d \phi}{d \tau} \Big)^2 = 2 \xi a(\tau)^2 \simeq \frac{2 \xi}{H^2 (\epsilon - 1)^2 \tau^2} ~,
\ee
and it is straightforward to solve $\phi(\tau)$ to be
\be \label{phi_tau}
 \phi(\tau) \simeq \Big( \frac{1}{\phi_i} \pm \frac{\sqrt{2\xi}}{H (1 - \epsilon) \sqrt{\lambda}} \ln\frac{\tau}{\tau_i} \Big)^{-1} ~,
\ee
where $\phi_i$ is the field value at the conformal time $\tau_i$ that is set to the beginning moment of inflation. Note that we also adopt the approximation that $H$ is regarded as a constant when we solve for $\phi(\tau)$ in \eqref{phi_tau}, this is reasonable as $H$ varies slowly in the quasi-de Sitter expansion. The solution \eqref{phi_tau} is also confirmed by the numerical results in Figs. \ref{fig:Hubb&V} and \ref{fig:phi}. A more rigorous treatment is performed in the Appendix, which also shows the validity of the approximated solution \eqref{phi_tau}. 
Since the term $\ln(\tau / \tau_i)$ is always negative during inflation, the ``+" sign refers to an increasing $\phi$ associated with the case of IR DBI, while ``-" represents a decreasing solution corresponding to the case of UV DBI. As the model UV DBI suffers from over large non-Gaussianity \cite{Baumann:2006cd, Lidsey:2006ia}, in the present study we focus on the case of IR DBI. Therefore, we stick to the increasing solution
\be \label{phi_Solution}
 \phi(\tau) \simeq \Big( \frac{1}{\phi_i} + \frac{\sqrt{2\xi}}{H (1 - \epsilon) \sqrt{\lambda}} \ln\frac{\tau}{\tau_i} \Big)^{-1} ~.
\ee
Moreover, the conformal time can be expressed in terms of the inflaton field $\phi$ as
\be \label{tau}
\tau = \tau_i \exp \Big[ \frac{H (1 - \epsilon) \sqrt{\lambda} }{\sqrt{2\xi}} (\phi^{-1} - \phi_i^{-1}) \Big] ~.
\ee

The next step is to involve the oscillating feature $\cos(2 k_*\tau)$ into the sound speed squared. Since the amplitude of oscillation can be quite small (i.e. $\xi < 1/4$) from \eqref{ssr}, the natural consideration is to regard this oscillating term as a consequence of small classical perturbation of the warp factor $f(\phi)$ or inflaton's evolution $\phi(\tau)$ or both of them in the above step. As the dynamical evolution of $\phi(\tau)$ must satisfy Eq. \eqref{EOM_DBI}, which is quite difficult to be solved analytically, we suggest to phenomenologically modify $f(\phi)$ to embed the additional oscillating term into the sound speed squared. We would like to clarify that, the search for such a solution of the warp factor from some exact string compactification is beyond the scope of this article, and leave it as an open question for future study.

\begin{figure}[h]
	\centering
	\includegraphics[width=0.9 \linewidth]{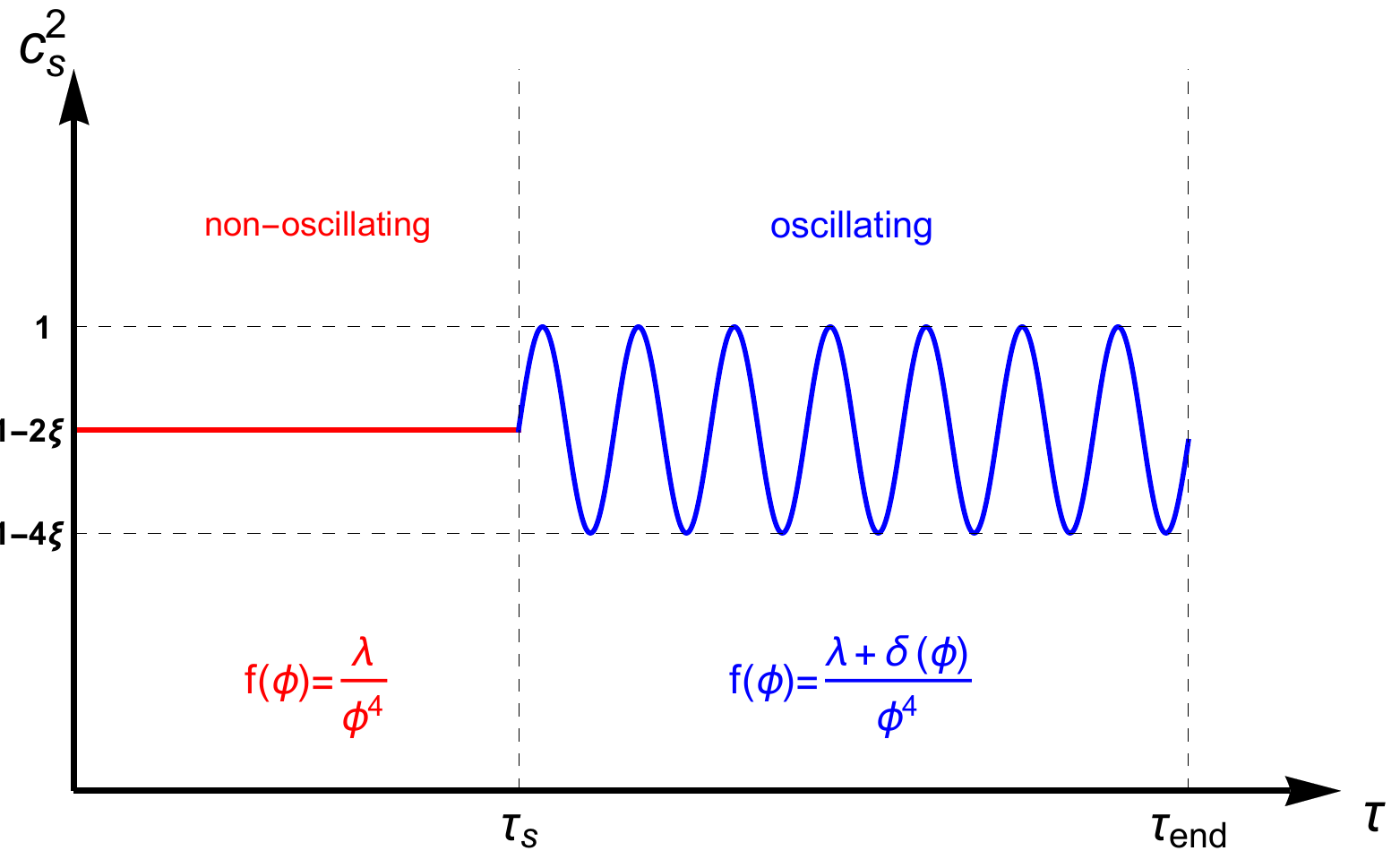}
	\caption{The schematic diagram of the sound speed squared $c_s^2$ of SSR within DBI inflation. The red straight line refers to the nonoscillating stage before the beginning moment of SSR $\tau_s$, while the blue curve represents the oscillating stage required by SSR from $\tau_s$ to a moment near the end of inflation $\tau_\text{end}$. The warp factor takes the standard form $f(\phi) = \lambda/\phi^4$ in the nonoscillating stage, and then deforms to $f(\phi) = (\lambda + \delta(\phi))/\phi^4$ in the oscillating stage.}
	\label{fig:cs}
\end{figure}

In our perturbative approach, the evolution of $\phi(\tau)$ in \eqref{phi_Solution} remains almost unchanged. Thus, the small deviation of the warp factor $f(\phi)$ is written as
\be \label{fphi}
f(\phi) = \frac{\lambda + \delta(\phi)}{\phi^4} ~.
\ee
Note that, the classical perturbative function $\delta(\phi)$ can be solved from the matching conditions \eqref{Matching_1} and \eqref{Matching_2}, and the solution of $\phi$ in \eqref{phi_Solution}, which yield
\begin{align} \label{delta}
 \delta(\phi) &= - \lambda C(\phi) ~, \nn \\ 
 C(\phi) &\equiv \cos \Big\{ 2 k_* \tau_s \exp \big[ \frac{H (1 - \epsilon) \sqrt{\lambda} }{\sqrt{2\xi}} (\frac{1}{\phi} - \frac{1}{\phi_s}) \big] \Big\} ~,
\end{align}
where $\phi_s = \phi(\tau_s)$ is the field value at the beginning moment of the oscillating stage. Accordingly, the warp factor becomes
\be \label{Warp}
 f(\phi) = \frac{\lambda \Big[ 1 - \Theta(\phi - \phi_s) C(\phi) \Big] }{\phi^4} ~,
\ee
where the Heaviside step function $\Theta(\phi - \phi_s)$ is introduced to simply represent the beginning moment of sound speed oscillation in \eqref{ssr}. Thus, our model involves the nonoscillating stage and the oscillating stage regarding the sound speed squared $c_s^2$ for the inflaton field during inflation, which is shown in Fig.~\ref{fig:cs} schematically. Before the beginning time $\tau_s$ of sound speed oscillation, the sound speed squared is fixed to $c_s^2 = 1 - 2 \xi$ with an AdS warp factor $f(\phi) = \lambda/\phi^4$; when entering the oscillating stage $\tau_s < \tau < \tau_\text{end}$, $c_s^2$ oscillates periodically between $1-4\xi$ and $1$ with the deformed warp factor $f(\phi) = (\lambda + \delta(\phi))/\phi^4$. We stress that the time evolution of inflaton $\phi(\tau)$ over the whole stages including the nonoscillating stage and the oscillating stage has the unique solution \eqref{phi_Solution}. We also notice that, as opposed to the original SSR mechanism \cite{Cai:2018tuh}, where the sound speed is assumed to start oscillating from $c_s^2 = 1$ to $c_s^2 = 1 -2 \xi [ 1-\cos(2k_*\tau) ]$ smoothly, the conjunction of the sound speed in our model here is slightly different, i.e., converting from $c_s^2 = 1 - 2 \xi$ to $c_s^2 = 1 -2 \xi [ 1-\cos(2k_*\tau) ]$, which is continuous but not smooth (the first time derivative of sound speed is not continuous). However, our analysis in Sec. \ref{sec:viability} shows that the narrow resonance effect in the SSR mechanism is barely influenced by this nonsmoothing conjunction of sound speed at the beginning moment of the oscillating stage.

\begin{figure}[h]
	\centering
	\includegraphics[width=0.9 \linewidth]{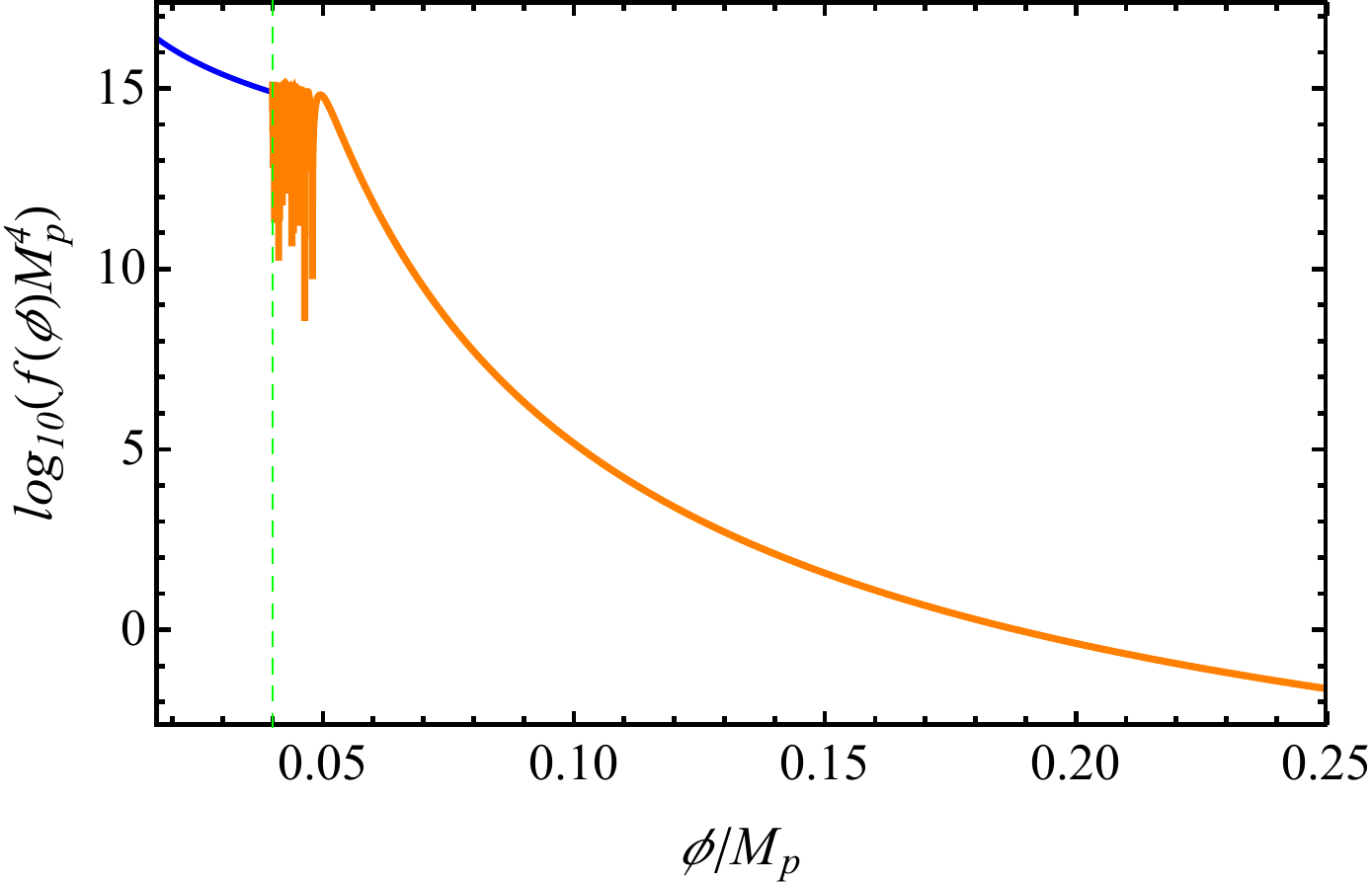}
	\caption{The warp factors in the nonoscillating stage $f(\phi) = \lambda/\phi^4$ (the blue curve) and in the oscillating stage $f(\phi) = (\lambda + \delta(\phi))/\phi^4$ (the orange curve). The green dashed line represents the beginning moment of the oscillating stage. The parameter values are chosen to be $\lambda = 2 \times 10^9$, $H_0 = 10^{-5} M_p$, $\xi = 0.1$ and $N_s = 21$.}
	\label{fig:warp}
\end{figure}

The warp factors in the nonoscillating stage and in the oscillating stage are shown in Fig.~\ref{fig:warp}, and in light of the observational bounds on the parameter space of our model discussed in Sec. \ref{sec:constriant}, we choose the values of parameters as $\lambda = 2 \times 10^9$, $H_0 = 10^{-5} M_p$, $\xi = 0.1$ and $N_s = 21$. One can read from the plot that the warp factors share the same power-law form of $f(\phi) \propto \phi^{-4}$ in both stages after moduling the oscillating feature. Also, we mention that the warp factor $f(\phi)$ starts to oscillate rapidly when it enters the oscillating phase and then behaves like $f(\phi) \propto \phi^{-4}$ near the end of inflation.

After that, we solve the Hubble parameter $H(\phi)$ and the potential $V(\phi)$ with the Hamilton-Jacobi formalism. Using the solution \eqref{phi_Solution}, Eq. \eqref{H_phi} reads
\be \label{H_1}
 H'(\phi) = - \sqrt{ \frac{\xi}{2 \lambda \big[ 1 - 2\xi ( 1 - \Theta(\phi - \phi_s) C(\phi) ) \big] } } \frac{\phi^2}{M_p^2} ~.
\ee

Although the above equation is quite complicated to get an exact analytical solution, one can still solve it in the nonoscillating stage by applying $c_s^2 = 1 - 2 \xi$, which yields
\be
 H'(\phi) = - \sqrt{ \frac{\xi}{2 \lambda (1 - 2 \xi)} } \frac{\phi^2}{M_p^2}  ~,
\ee
and thus, one obtains 
\be \label{H_Solution}
 H(\phi) = H_0 - \sqrt{ \frac{\xi}{2 \lambda (1 - 2 \xi)} } \frac{\phi^3}{3 M_p^2} ~,
\ee
where $H_0 = H(\phi_i) + \sqrt{ \frac{\xi}{2 \lambda (1 - 2 \xi)} } \frac{\phi_i^3}{3 M_p^2}$. Plugging \eqref{H_Solution}
into \eqref{V_phi} to obtain the approximate solution of the potential
\be \label{Potential}
 V(\phi) = 3 H_0^2 M_p^2 - \sqrt{ \frac{2 \xi}{\lambda (1 - 2 \xi)} } H_0 \phi^3 + \mathcal{O}(\phi^4)~.
\ee
As a long period of inflation can occur near the top of the potential, the first two terms in the potential \eqref{Potential} dominate.

\begin{figure}[h]
\centering
\includegraphics[width=0.9 \linewidth]{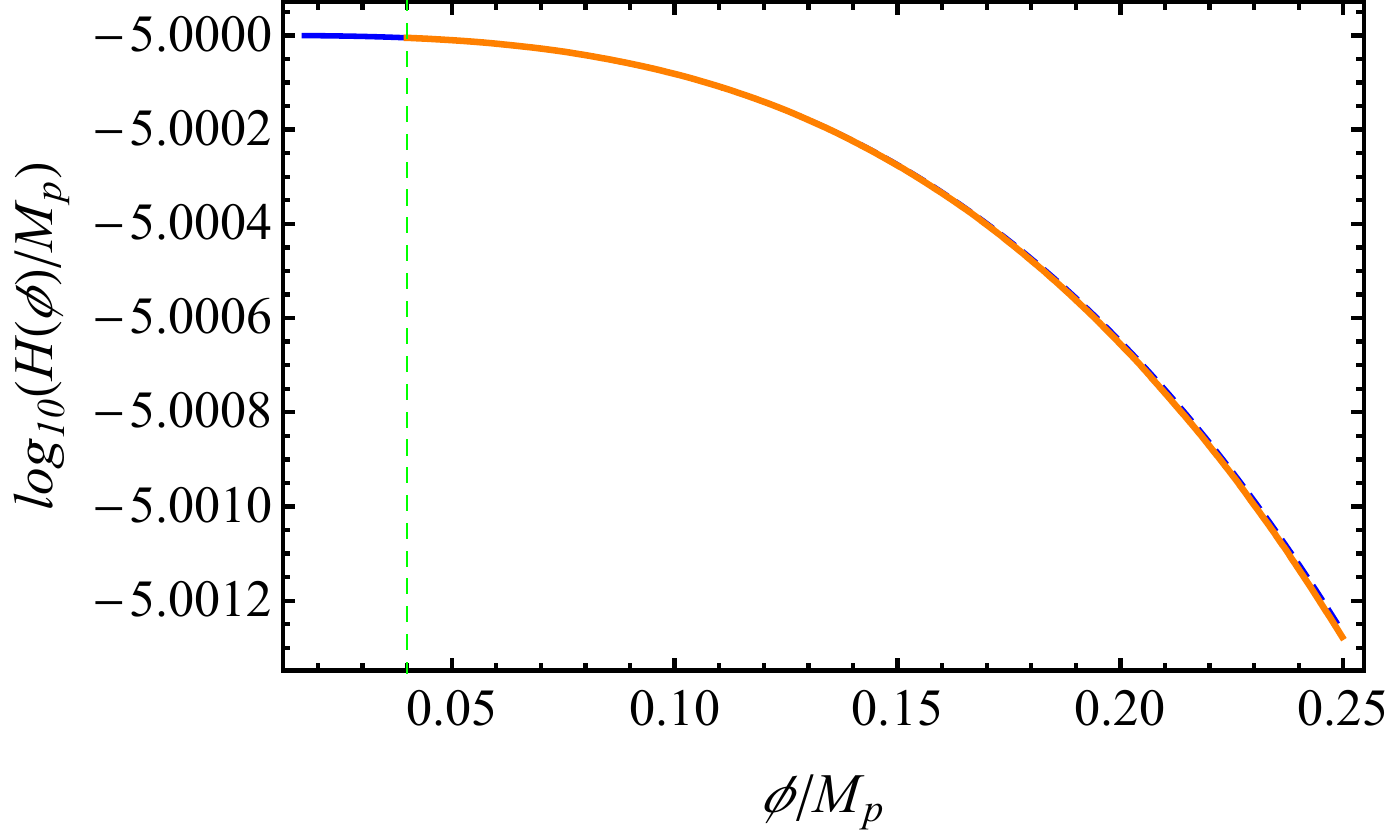}
\includegraphics[width=0.9 \linewidth]{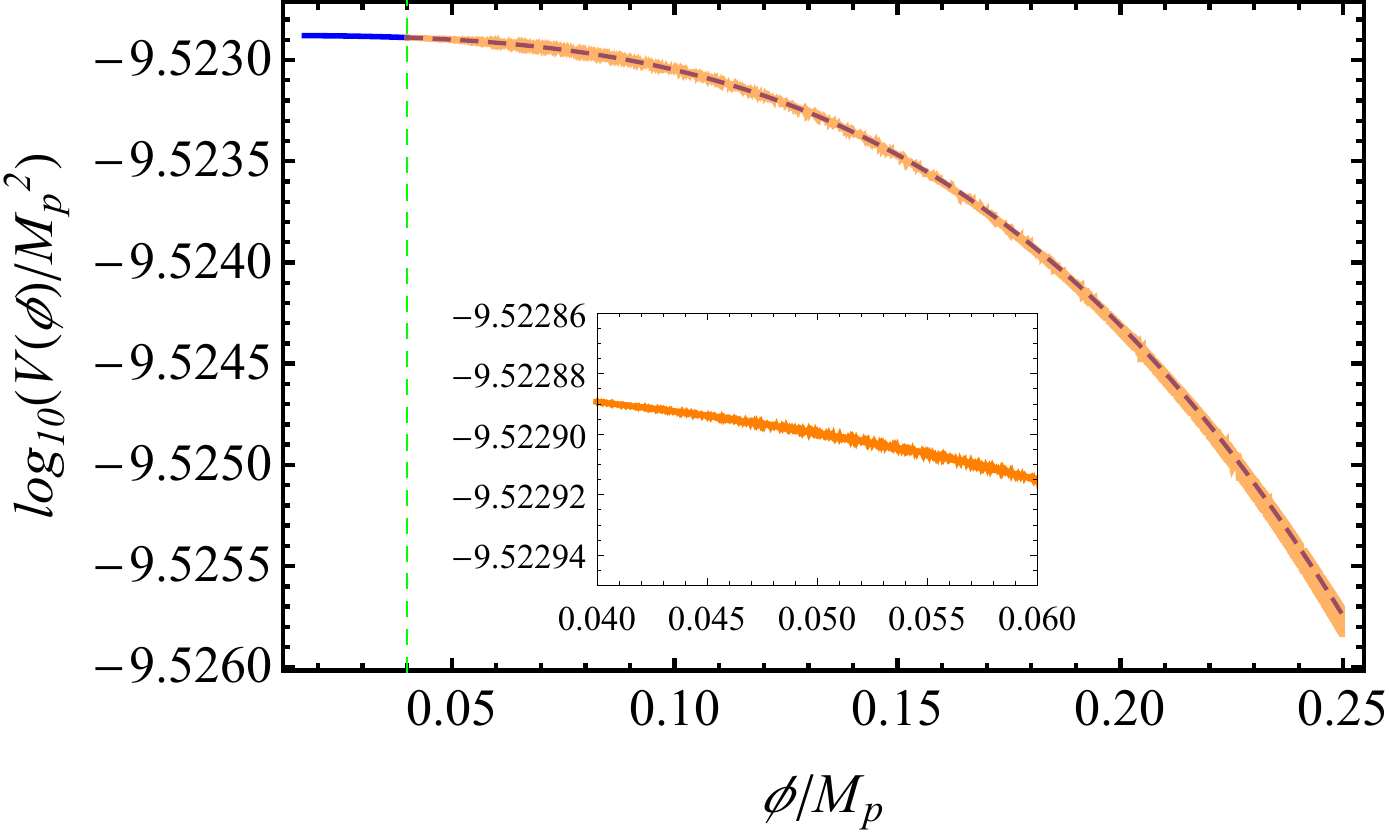}
\caption{The forms of the Hubble parameter and the inflaton's potential as functions of $\phi$ from the beginning $\phi_i$ to the end of inflation $\phi_\text{end}$. The green dashed lines refer to the beginning of the oscillating stage. The comparisons between the approximate results \eqref{H_Solution} and \eqref{Potential} and the numerical results in the oscillating stage are made by the blue dashed curves and the orange curves, respectively. The parameter values are chosen to be: $\lambda = 2 \times 10^9$, $H_0 = 10^{-5} M_p$, $\xi = 0.1$ and $N_s = 21$. }
\label{fig:Hubb&V}
\end{figure}

Moreover, due to the fact that throughout the whole analysis $\xi$ is a small quantity, the solutions of the Hubble parameter $H(\phi)$ in \eqref{H_Solution} and the potential $V(\phi)$ in \eqref{Potential} can approximately hold in both the nonoscillating and the oscillating stages, which have been confirmed by numerical analyses, see Fig.~\ref{fig:Hubb&V}. We integrate Eq. \eqref{H_1} numerically in the oscillating stage and then insert the corresponding result into \eqref{V_phi} to calculate the potential. One can see that, when inflaton occurs around the top of the potential in \eqref{Potential}, the Hubble parameter in \eqref{H_Solution} is nearly a constant $H \simeq H_0$. The inflaton's potential $V(\phi) \simeq 3 H_0^2 M_p^2$ is roughly a constant as well at the beginning of inflation. Therefore, the assumption of the quasi-de Sitter background made in the semianalytical calculation is reliable, which can also be read from the comparison with numerical estimations.

\section{Theoretical viability} \label{sec:viability}

According to the previous section, we have arrived at a concrete DBI realization for SSR with a specific warp factor in \eqref{Warp} and the inflaton's potential in \eqref{Potential}. Moreover, in order to test the theoretical viability of this reconstructed model, we in this section make the detailed investigations combined with relevant discussions. 

\subsection{The numerical analysis of inflaton evolution}

\begin{figure}[h]
	\centering
	\includegraphics[width=0.9 \linewidth]{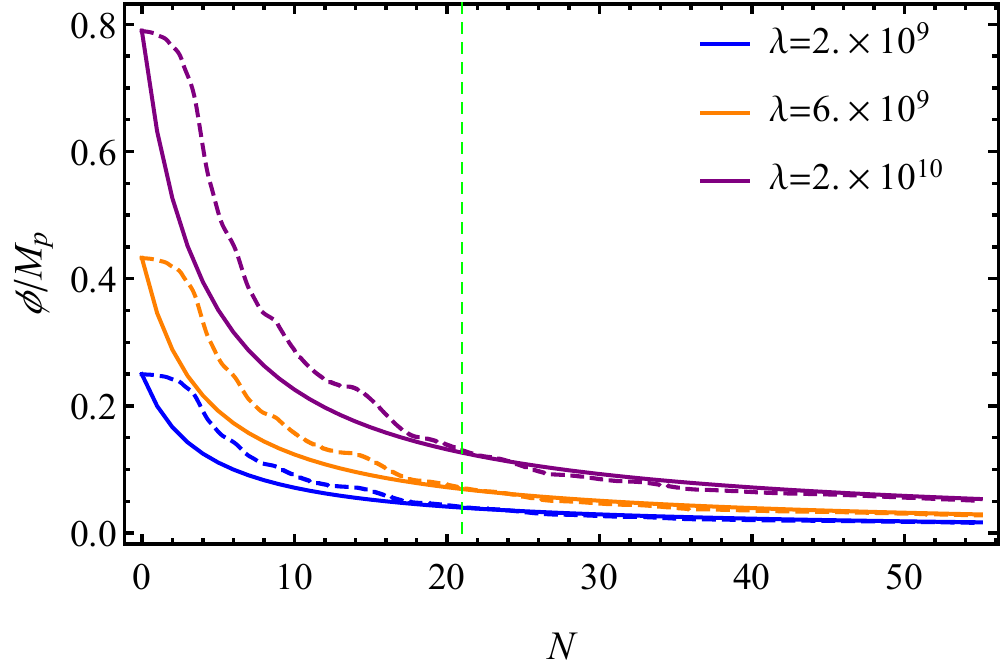}
	\caption{Numerical (the dashed curves) and semianalytical results (the solid curves) of the evolutions of the inflaton field $\phi$ with different values of $\lambda$. The green dashed line denotes the beginning of the oscillating stage. The parameter values are chosen to be: $H_0 = 10^{-5} M_p$, $\xi = 0.1$ and $N_s = 21$. }
	\label{fig:phi}
\end{figure}

For a set of different values for $\lambda$ (or equivalently, different values of the field value $\phi_\text{end}$ at the end of inflation through Eq. \eqref{phi_end}), the comparisons between the numerical results and the semianalytical approximations of the evolutions of inflaton field $\phi$ along with the $e$-folding number $N$ \eqref{phi_Solution} are presented in Fig.~\ref{fig:phi}. In the numerical calculations, the expressions of the warp factors in the nonoscillating stage and the oscillating stage are regarded as the inputs of our model. After that, the numerical analyses for the evolutions of $\phi$ are accomplished by virtue of the matching conditions \eqref{Matching_1} and \eqref{Matching_2}, and also the background equation \eqref{H_phi}, for the phenomenological sound speed squared $c_s^2$. One can read from Fig.~\ref{fig:phi} that our semianalytical solution in \eqref{phi_Solution} is good enough to describe the evolution of the inflaton field in both the nonoscillating and oscillating stages. Note that, the $e$-folding number here is defined as $N \equiv \ln a(\tau_\text{end}) /a(\tau)$ which measures the number of $e$-folds from the moment $\tau$ to the end of inflation $\tau_\text{end}$. Hence, $N=0$ represents the end of inflation and larger $N$ corresponds to the earlier time during inflation. These results also demonstrate the validity of our perturbative approach that was used to reconstruct such a concrete DBI realization for SSR.

\subsection{PBH mass function}

\begin{figure}[h]
	\centering
	\includegraphics[width=0.9 \linewidth]{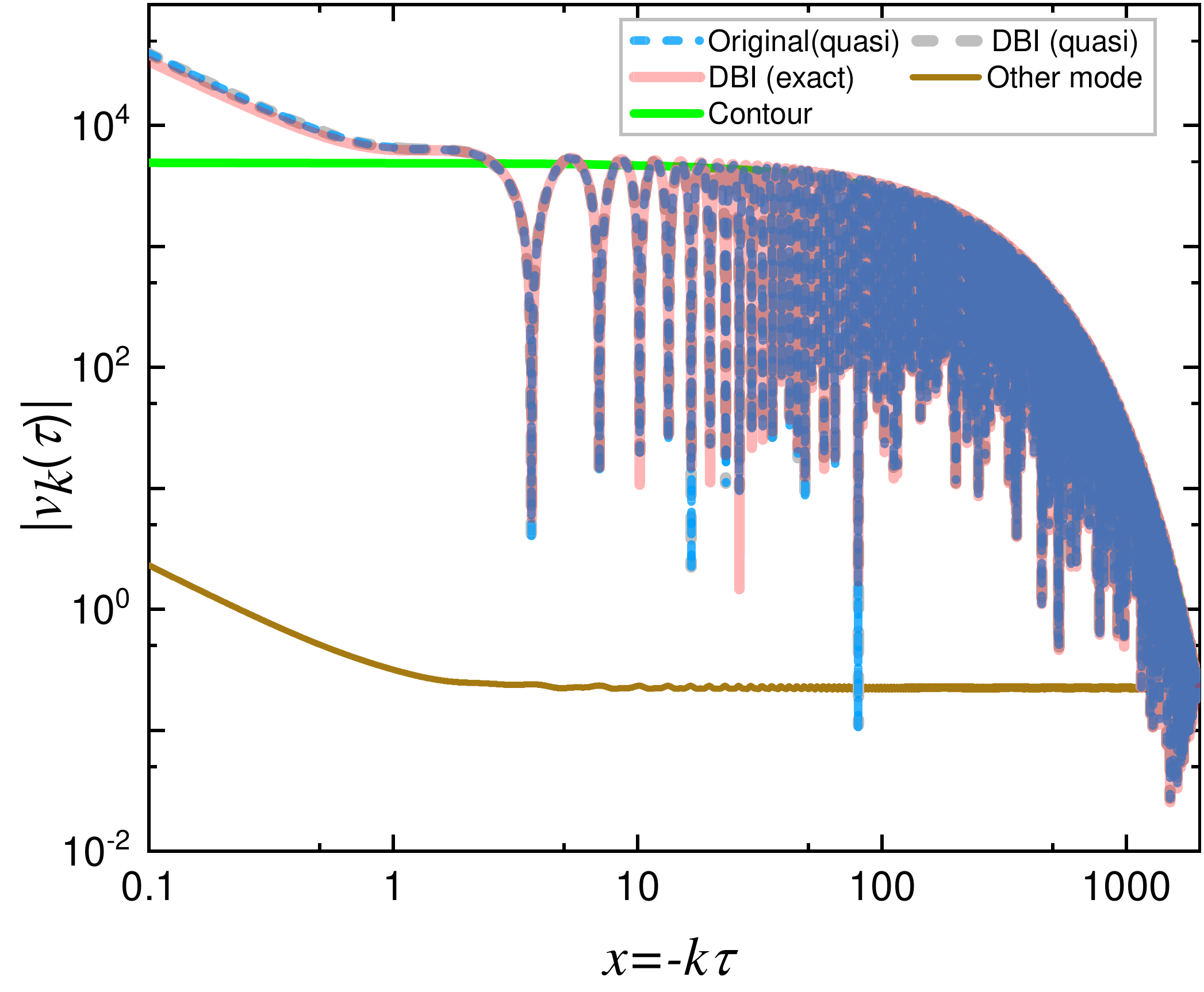}
	\includegraphics[width=0.9 \linewidth]{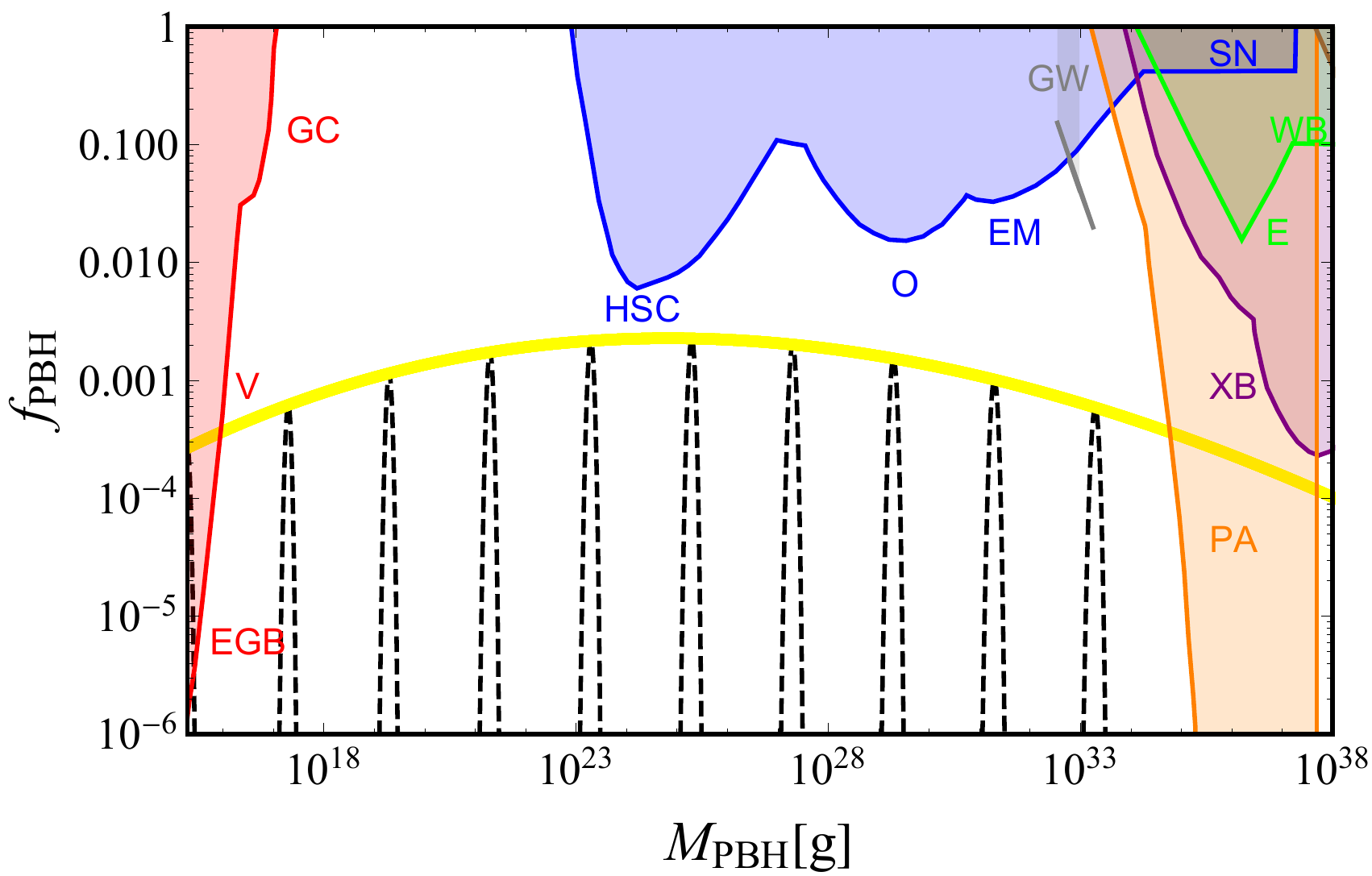}
	\caption{Upper: Mode functions derived from the original SSR and the DBI SSR models, respectively. The evolutions of the resonating modes under the quasi-de Sitter approximation for the nonsmooth conjunction of DBI SSR (the grey dashed curve) and for the smooth conjunction of original SSR (the blue dashed curve) match well, and the exact numerical results in DBI SSR (the red solid curve) also match with them to a certain extent. The brown solid line denotes a non-resonating mode $k \neq k_*$. The growth of mode function is estimated as $|v_{k_c}(\tau)| \propto \exp(\xi k_* \tau/2)$ (the green solid curve). Lower: The mass spectrum of PBH $f_\text{PBH}$ in the DBI SSR, for different values of $k_*$. The colored shadow areas refer to various astronomical constraints displayed in Fig. 1 of \cite{Carr:2020xqk}: constraints from evaporations (red), lensing (blue), gravitational waves (grey), dynamical effects (green), accretion (purple) and CMB distortions (orange). 
	}
	\label{fig:mode}
\end{figure}

As we have mentioned in Sec. \ref{sec:realization}, the conjunction of the sound speed at the beginning moment of the oscillating stage in our model is continuous but not smooth in contrast to the original SSR mechanism. Thus, it is necessary to examine the possible influence of this nonsmooth conjunction of sound speed on the SSR phenomenology. Analogous to the treatments in \cite{Cai:2018tuh, Chen:2019zza}, we introduce a canonical variable $v \equiv z \zeta$ for the comoving curvature perturbation $\zeta$, where $z = \sqrt{2 \epsilon} M_p a/c_s$. The evolution of a Fourier mode of this variable $v_k(\tau)$ satisfies the Mukhanov-Sasaki equation $\frac{d^2 v_k}{d \tau^2} + \l(c_s^2 k^2 - \frac{1}{z} \frac{d^2 z}{d \tau^2} \r) v_k = 0$ \cite{Mukhanov:1988jd, Sasaki:1986hm}. We numerically solve this equation by setting the initial mode in the nonoscillating stage to the renormalized Bunch-Davies (BD) vacuum, i.e. $v_k(\tau) = e^{- i \sqrt{1 - 2 \xi} k \tau}/{\sqrt{2 \sqrt{1 - 2 \xi} k}}$. The results in the upper panel of Fig.~\ref{fig:mode} imply that in the quasi-de Sitter approximation, the evolutions of $v_k(\tau)$ in the resonant regime around the characteristic scale $k_*$ (the grey solid curve) in the DBI model match very well with the results of the original SSR (the blue dashed curve) \cite{Cai:2018tuh, Chen:2019zza}, as well as the exact numerical results of DBI SSR (the red solid curve), which can be solved out by combining the matching condition \eqref{Matching_1}, the background evolution \eqref{H_phi} and the warp factor \eqref{Warp}. The above comparisons indicate that SSR is dominated by the narrow resonance effect in the oscillating stage, and insensitive to the nonsmooth conjunction of sound speed at the beginning moment of the oscillation. Consequently, the fraction of PBH against the total dark matter density $f_\text{PBH} \equiv \Omega_\text{PBH} / \Omega_\text{DM}$, where $\Omega_\text{PBH}$ and $\Omega_\text{DM}$ are the corresponding normalized energy densities of PBHs and dark matter at the present time, is expected to be the same as the original SSR \cite{Cai:2018tuh, Chen:2019zza}, which is presented in the lower panel of Fig.~\ref{fig:mode}.

Additionally, regarding the general speed limit on DBI models \eqref{Speed_Limit}, we derive the bound on the amplitude of sound speed oscillation $\xi$ in SSR by the background solution \eqref{phi_Solution} and the warp factor \eqref{fphi}, i.e.,
\be \label{xi}
 \xi \leq \frac{\lambda}{2 (\lambda + \delta(\phi))} ~,
\ee
and thus, $\xi < \frac12$ is required in the nonoscillating stage $\tau < \tau_s$ and $\xi \leq \frac14$ in the oscillating stage $\tau_s < \tau < \tau_\text{end}$. Accordingly, in the SSR mechanism where $\xi < \frac14$ is required for the positivity of $c_s^2$ \cite{Cai:2018tuh, Cai:2019jah, Chen:2019zza}, the speed limit \eqref{Speed_Limit} is always satisfied. In the nonoscillating stage, Eq. \eqref{xi} also implies that a small amplitude of $\xi$ corresponds to the nonrelativistic motion of DBI inflaton. In fact, from the string theory perspective, the velocity of a brane in the oscillating stage may be estimated as $v_p = \sqrt{f(\phi)} \dot{\phi} = \sqrt{2 \xi (1+ {\delta}/{\lambda})}$, which evolves between $0$ and $2 \sqrt{\xi}$.

\subsection{Parameter space}

Furthermore, we would like to comment that there are three categories of model parameters. The first class is the microscopic parameter $\lambda$, which appears in the warp factor. Making use of experimental bounds on the amplitude of primordial density perturbation, we derive $\lambda \gtrsim 1.3 \times 10^9$ for $\xi = 0.1$, which can be seen from the constraint \eqref{Lambda} and the discussions in Sec. \ref{sec:powerspectra}. The second class is the parameter of the inflationary background, namely, $H_0 = 10^{-5} M_p$ is applied in this work. The values of $H_0$ and $\lambda$ restrict the slow-roll parameter \eqref{slowroll1} to be around $\tilde{\epsilon} \simeq 0.001$, which will be shown in Sec. \ref{sec:powerspectra}. The last class concerns the SSR parameters. For the characteristic scale $k_*$, we fix $- k_* \tau_s \simeq e^{\Delta N_*} = 200$ or $\Delta N_* \simeq 5.3$, which corresponds to the $e$-folding number from $\tau_s$ to Hubble exit of $k_*$ \cite{Cai:2018tuh, Cai:2019jah, Chen:2019zza}. It is straightforward to convert the conformal time into the numbers of $e$-folds: $- k_* \tau \simeq - k_* \tau_s e^{N - N_s}$, where $N$ is the $e$-folding number from $\tau$ to the end of inflation $\tau_\text{end}$, and $N_s$ is the $e$-folding number from $\tau_s$ to $\tau_\text{end}$. Using the horizon-mass approximation for the PBH mass at the reentry of Hubble radius, one can relate the $N_s$ to the PBH mass $M_*$ as $N_s = \ln (\tau_s / \tau_\text{end}) \simeq 40 + \frac12 \ln (M_* / M_\odot)$ \cite{Sasaki:2018dmp}, where $M_*$ is the horizon mass at the scale $k_*$ and $M_\odot$ is the solar mass. For instance, a PBH with mass $10^{17}$ g corresponds to the $e$-folding number $N_s \simeq 21$.

Accordingly, the evolution of $\phi$ with the numbers of $e$-folds can be expressed roughly as $\phi(N) \simeq \l[ \phi_i^{-1} + \frac{\sqrt{2\xi}}{H (1 - \epsilon) \sqrt{\lambda}} (N - N_\text{end}) \r]^{-1}$ from the solution \eqref{phi_Solution}, where $N$ is the $e$-folding number from $\tau$ to the end of inflation $\tau_\text{end}$, and then the approximate field value at the beginning of the oscillating phase is calculated as, for example, $\phi_s = \phi(\tau_s) = \phi(N_s = 21) \simeq 0.04 M_p$ by setting $\lambda = 2 \times 10^9$, $N_s = 21$ and $N_\text{end} = 55$.

\subsection{Adiabaticity analysis}

One may be concerned about whether the adiabatic condition of the Mukhanov-Sasaki equation $\frac{d^2 v_k}{d \tau^2} + \l(c_s^2 k^2 - \frac{1}{z} \frac{d^2 z}{d \tau^2} \r) v_k = 0$ is violated or not when the time-oscillating sound speed \eqref{ssr} is introduced in the SSR mechanism, i.e., we need to examine the adiabatic condition $|\omega' / \omega^2 | \ll 1$ holds or not during the oscillating stage $\tau_s < \tau < \tau_\text{end}$, where $\omega^2 \equiv c_s^2 k^2 - \frac{1}{z} \frac{d^2 z}{d \tau^2} $. The following analysis demonstrates that the evolution of the curvature perturbation $\zeta_k$ is not affected by the violation of adiabatic condition of $v_k$. The key point is that the Mukhanov-Sasaki variable $v_k$ is not a true physical quantity, and the apparent violation stems from the definition $v_k \equiv \sqrt{2 \epsilon}  M_\text{pl} a / c_s \zeta_k$ involving the nontrivial sound speed $c_s$ which oscillates rapidly at the early stage of the oscillation, while the real physical mode $\zeta_k$ behaves well (similar to the BD vacuum in the nonresonant region). And we also show that the final power spectra in the nonresonant region $k < k_*$ are barely affected by the nontrivial sound speed \eqref{ssr} after Hubble crossing. 

\begin{figure}[h]
	\centering
	\includegraphics[width=0.48 \linewidth]{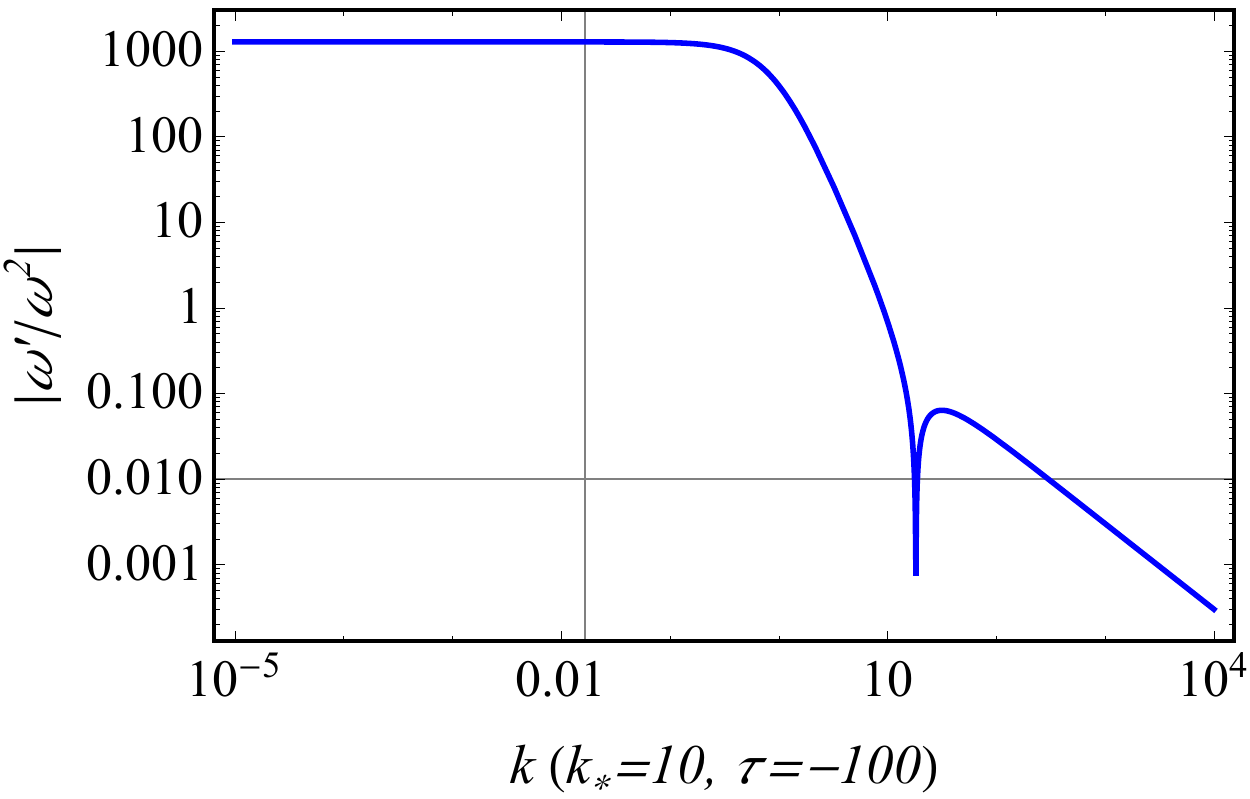}
	\includegraphics[width=0.48 \linewidth]{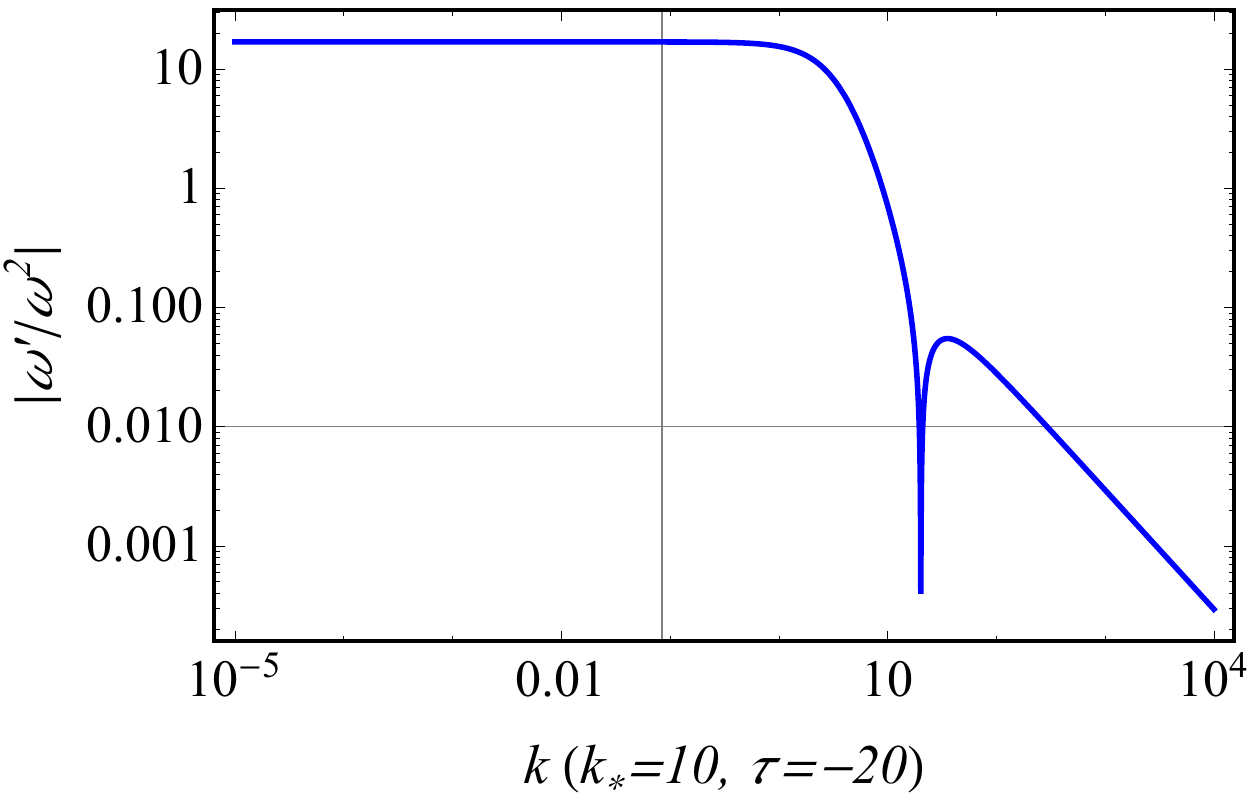}
	\includegraphics[width=0.48 \linewidth]{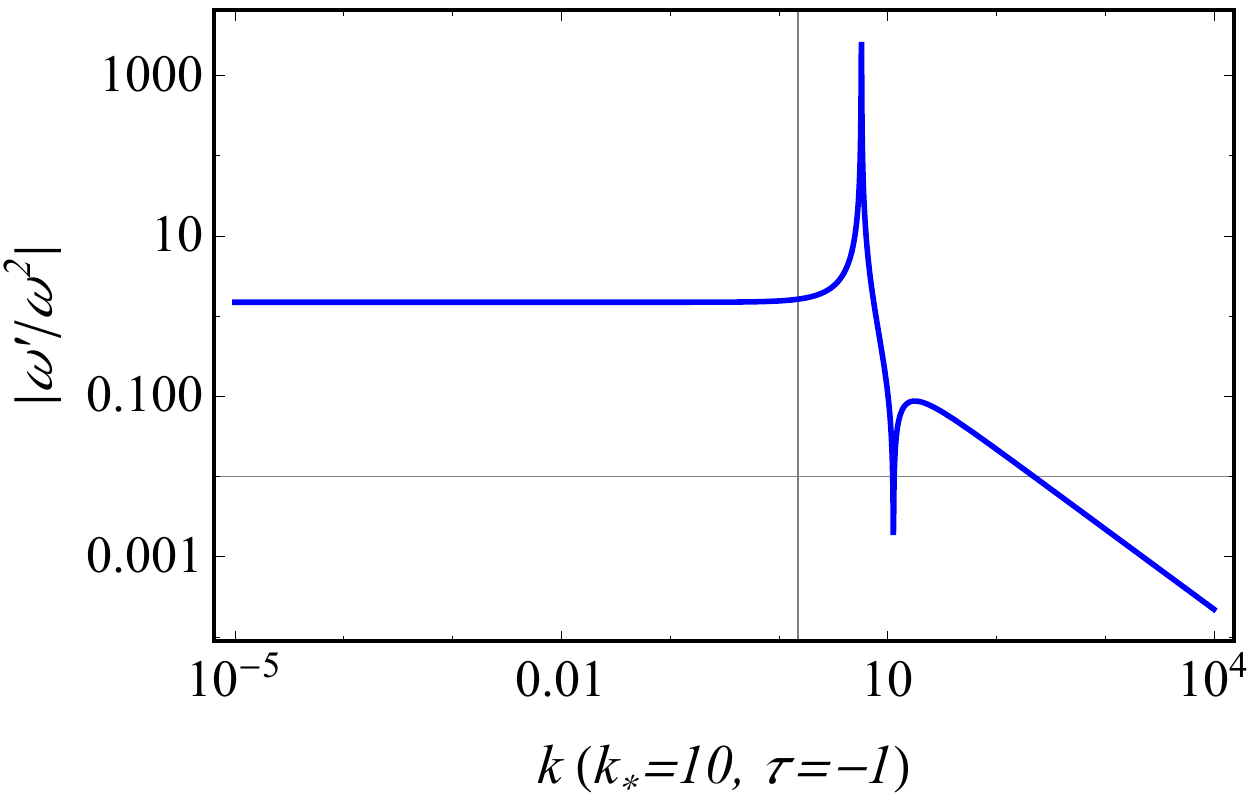}
	\includegraphics[width=0.48 \linewidth]{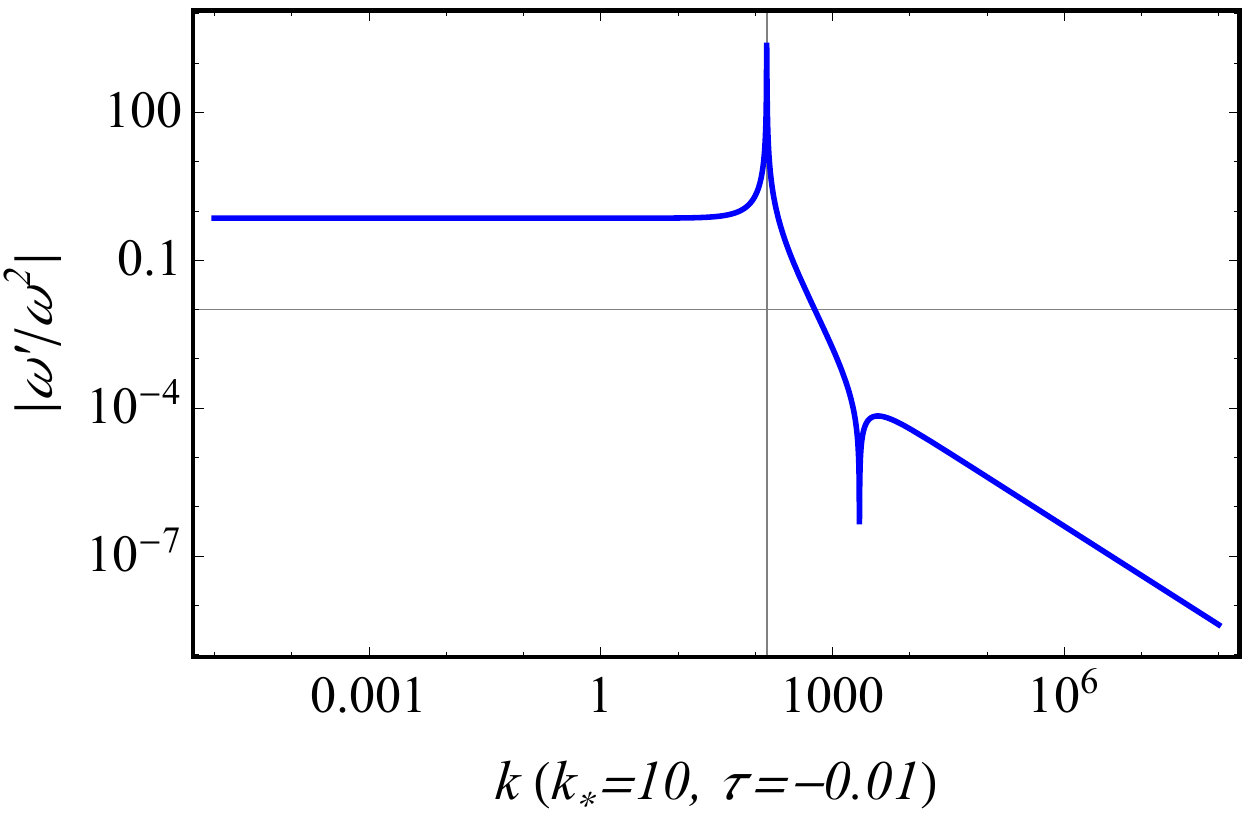}
	\caption{$\Big| \frac{\omega'}{\omega^2} \Big|$ as a function of $k$ at different conformal times $\tau = ( -100, -20, -1, -0.01 )$ in the SSR mechanism (de Sitter approximation for the background evolution) with the time-oscillating sound speed \eqref{ssr}. The vertical lines at $k = - \sqrt{2} / (\tau c_s)$ and the horizontal lines at the small value $0.01$. We fix $\xi = 0.1$ and $k_* = 10$.}
	\label{fig:adiabatic}
\end{figure}

In the usual slow-roll case, the sound speed is equal to the speed of light $c_s = 1$ during the whole inflationary expansion, and we yield $|\omega' / \omega^2 | = \frac{2}{ (k^2 - 2 / \tau^2)^{3/2} (- \tau)^3}$. Obviously, the adiabatic condition is always violated at the Hubble crossing (around the  singularity $k = - \sqrt{2} / \tau$), which is consistent with the usual statement that the classical perturbations generate at the horizon crossing. In the SSR mechanism, this adiabatic condition is indeed violated, as shown in Fig. \ref{fig:adiabatic}. It is clear that the adiabatic condition of $v_k$ is badly violated in the nonresonant region $k < k_*$, i.e., $|\omega' / \omega^2 | \ll 1$ does not hold even at the super-Hubble scale for some small $k$ modes, which seems to mean that the long-wave perturbations would still evolve after Hubble crossing in the SSR mechanism.

\begin{figure}
	\centering
	\includegraphics[width=0.8 \linewidth]{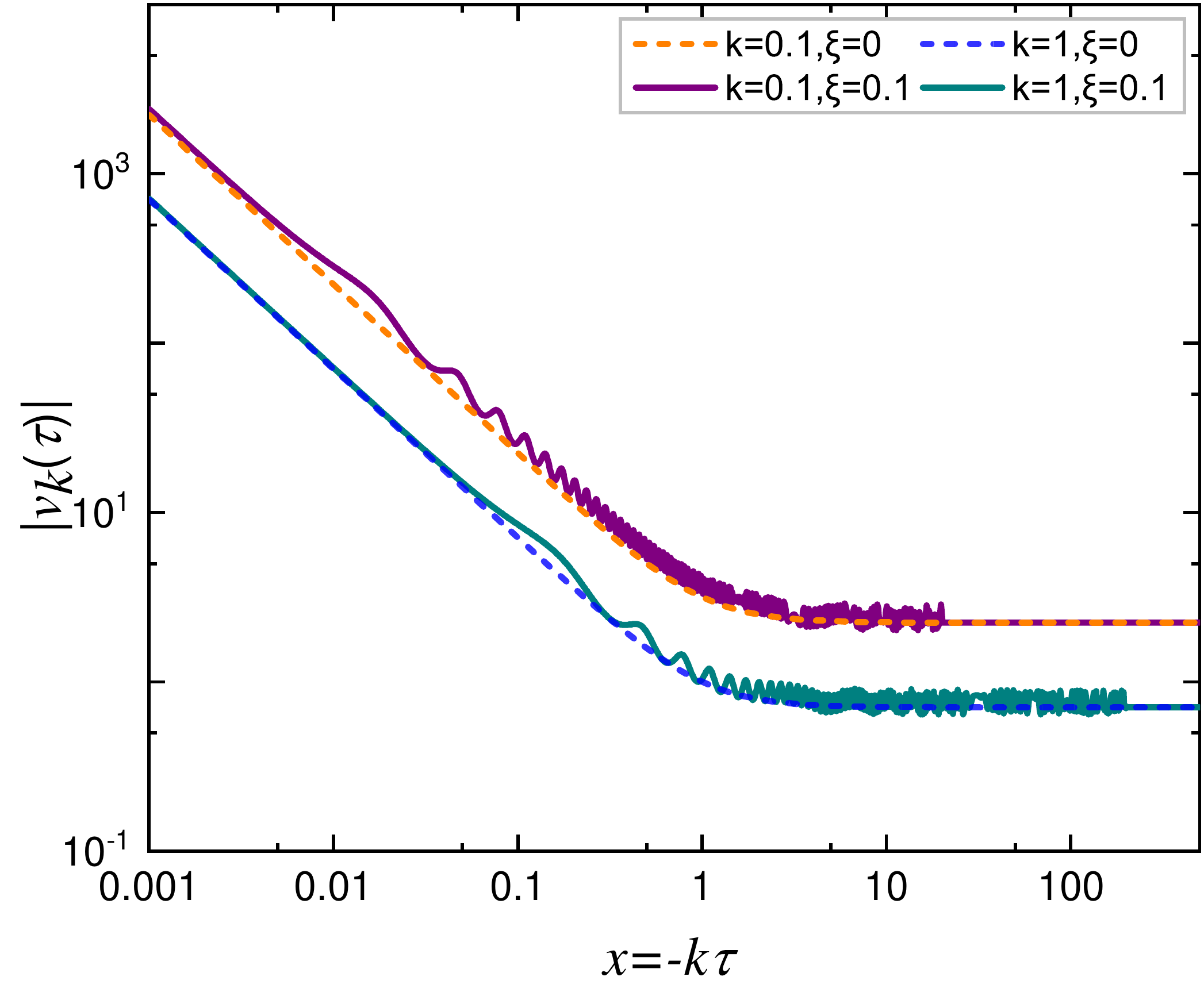}
	\includegraphics[width=0.8 \linewidth]{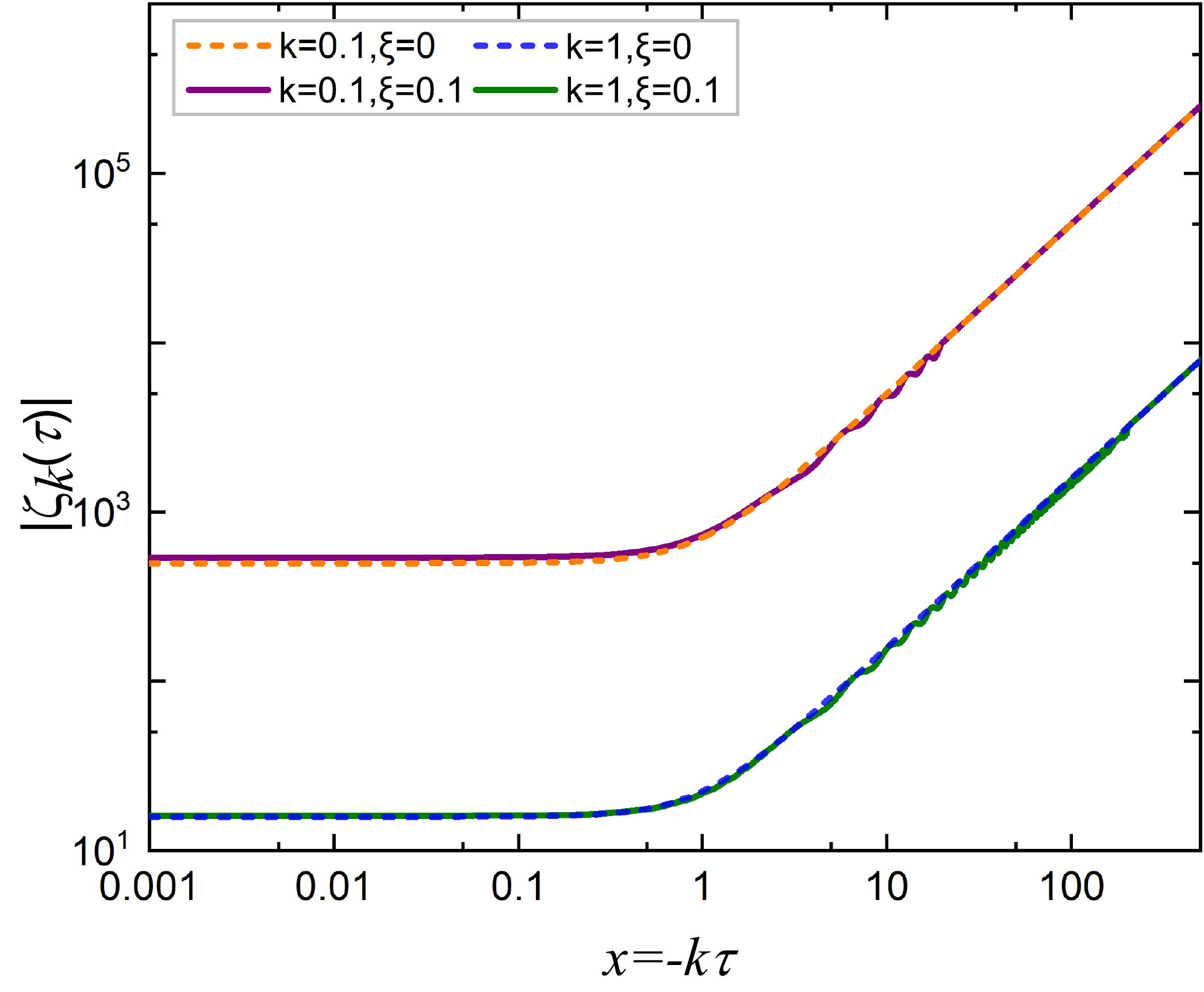}
	\caption{The comparison of the evolutions of mode functions $v_k$ and $\zeta_k$	in the nonresonant region $k (= 0.1, 1) < k_* (= 10)$ between the SSR mechanism ($\xi = 0.1$) and the BD vacua ($\xi = 0$). The solid lines represent the evolutions of $v_k$ and $\zeta_k$ in the de Sitter background in the SSR mechanism \cite{Cai:2018tuh, Chen:2019zza}, both of which match well with that of the BD vacua that are depicted by the dashed lines. }
	\label{fig:zeta_ad}
\end{figure}

It is not surprising that the evolution of $v_k$ violates the adiabatic condition, the major reason is that the time-oscillating $c_s$ is introduced in the definition of $v_k$. However, the real physical quantity $\zeta_k$ always behaves well, freezes after Hubble crossing. Figure \ref{fig:zeta_ad} shows that the evolution of $\zeta_k$ oscillates inside the Hubble radius due to the Mathieu solution of $v_k$ in the SSR mechanism \cite{Cai:2018tuh, Chen:2019zza}, and stops at Hubble crossing, which matches well with the evolutions of BD vacua ($\xi = 0$) which are represented by the dashed lines. The evolutions of $v_k$ are also similar to the BD modes at the sub- and super-Hubble scales. So, the violation of the adiabatic condition of $v_k$ does not affect the evolution of the curvature perturbation $\zeta_k$ at the super-Hubble scales in the SSR mechanism. However, the violation of the adiabatic condition seems to be problematic when we embed the SSR mechanism into an EFT framework \cite{Achucarro:2012yr}. Let us stress that the goal of this paper is to present a phenomenological realization of the SSR mechanism in the context of single field DBI inflation, which is a preliminary investigation on the phenomenological realization of SSR. In this sense, embeding the SSR mechanism into a UV completion theory is still a compelling problem in the follow-up study.

\section{Constraints} \label{sec:constriant}

In the previous section, we have performed a perturbative approach to achieve SSR in the context of DBI inflation with a specific warp factor \eqref{Warp} and the inflaton's potential \eqref{Potential}. It is well known that, for noncanonical inflation models, a nontrivial sound speed of inflaton can yield observable effects that are distinct from that of the regular model of canonical slow-roll inflation, e.g., a modified inflationary consistency relation and potentially large primordial non-Gaussianities \cite{Silverstein:2003hf, Peiris:2007gz, Bean:2007hc, Lidsey:2006ia}. In the following, we shall discuss these observables and their constraints on our model.

\subsection{The number of $e$-folds} \label{sec:efold}

In order to study the power spectra generated in DBI inflation, we introduce the following set of slow-roll parameters: 
\bl \label{slowrollabc}
\epsilon \equiv - \frac{\dot{H}}{H^2} ~,~ \eta \equiv \frac{\dot{\epsilon}}{\epsilon H} ~,~ \kappa \equiv \frac{\dot{c}_s}{H c_s} ~,
\el
which measure the variations of the Hubble parameter $H$, the first slow-roll parameter $\epsilon$, the sound speed $c_s$, with respect to each Hubble time, respectively. After that, we also introduce the following redefined slow-roll parameters in the context of the Hamilton-Jacobi formalism that we have used in the previous section,
\bl
 \tilde{\epsilon} &\equiv \frac{2 M_p^2}{\gamma} \Big( \frac{H'(\phi)}{H(\phi)} \Big)^2 ~, \label{slowroll1} \\
 \tilde{\eta} &\equiv \frac{2 M_p^2}{\gamma} \frac{H''(\phi)}{H(\phi)} ~, \label{slowroll2} \\
 \tilde{\kappa} &\equiv \frac{2 M_p^2}{\gamma} \frac{H'(\phi)}{H(\phi)} \frac{\gamma'(\phi)}{\gamma(\phi)} ~. \label{slowroll3}
\el
Using Eqs. \eqref{Lorentz_Factor} and \eqref{H_phi}, we can relate the above two sets of slow-roll parameters as follows, \cite{Bean:2007hc, Peiris:2007gz}
\bl
 \tilde{\epsilon} = \epsilon ~,~ \tilde{\eta} = 2 \epsilon - \eta - \kappa ~,~ \tilde{\kappa} = \kappa ~.
\el
In the nonrelativistic limit ($\gamma \rightarrow 1$), the slow parameters \eqref{slowroll1}, \eqref{slowroll2} and \eqref{slowroll3} relate to the usual slow-roll parameters $\epsilon_\text{sr} \equiv \frac{M_p^2}{2} \l( \frac{V'(\phi)}{V(\phi)} \r)^2$ and $\eta_\text{sr} \equiv M_p^2 \frac{V''(\phi)}{V(\phi)}$: $\epsilon \rightarrow \epsilon_\text{sr}$, $\eta \rightarrow \eta_\text{sr} - \epsilon_\text{sr}$ and $\kappa \simeq 0$.

For one thing, we follow \cite{Bean:2007eh} to check whether our model can produce a sufficiently long duration of inflationary expansion. Afterwards, we compare the observable predictions with the latest experimental data to narrow down the parameter space. In our case, the model belongs to the IR-type DBI inflation, and a long period of inflation can take place near the top of the potential. A quantitative check can be made to ensure the sufficient $e$-folding number from the time CMB quadruple exits the horizon to the end of inflation, i.e., $N_\text{cmb} \in [50, 60]$ \cite{Ade:2015lrj, Akrami:2018odb}. Using the solution of $\phi(\tau)$ \eqref{phi_Solution}, one obtains the total number of $e$-folds from the start of inflation to the end of inflation
\bl \label{efold}
 N_\text{end} \equiv \ln \frac{a_\text{end}}{a_i} =& -\int_{\phi_\text{end}}^{\phi_i} \frac{H(\phi)}{\dot{\phi}} d\phi \nn
 \\ \simeq& \frac{H_0 (1 - \epsilon) \sqrt{\lambda}}{\sqrt{2\xi}} ( \phi_i^{-1} - \phi_\text{end}^{-1} ) ~,
\el
where $\phi_\text{end} = \phi(\tau_\text{end})$ is the field value of inflaton at the end of inflation. The observational constraint then gives $N_\text{end} \geq N_\text{cmb}$. We also notice that the total $e$-folding number $N_\text{end}$ can also be derived directly from the solution of $\phi(\tau)$ in \eqref{phi_Solution} by considering $\tau_\text{end} / \tau_i \simeq e^{- N_\text{end}}$. One direct constraint from the $e$-folding number is the lower bounds on the inverse field range $( \phi_i^{-1} - \phi_\text{end}^{-1} )$ as follows,
\be \label{Field_Range1}
 \big( \phi_i^{-1} - \phi_\text{end}^{-1} \big) \gtrsim 55 \frac{\sqrt{2\xi}}{H_0  (1 - \epsilon) \sqrt{\lambda}} ~.
\ee
Here, we have taken a conservative value for the number of $e$-folds with $N_\text{cmb} = 55$ \cite{Akrami:2018odb}.


In the usual situation, inflaton ends when the slow-roll parameter $\epsilon$ tends to unity. However, in most models of DBI inflation $\epsilon$ remains less than 1, and inflation ends for different reasons depending on the underlying fundamental physics \cite{Bean:2007hc}. 
%
%
Despite the underlying fundamental theory, in our model inflaton ends subject to the condition of the second slow-roll parameter $|\eta| = 1$ \eqref{slowrollabc}. 

Using the Hubble parameter \eqref{H_1} and \eqref{H_Solution}, the slow-roll parameter $\epsilon$ \eqref{slowroll1} can be rewritten as
\be \label{SR}
\tilde{\epsilon} \simeq \frac{\xi}{\lambda} \frac{\phi^4}{H_0^2 M_p^2} ~.
\ee
Here it is shown that $\tilde{\epsilon} \propto \phi^4$ is quite small. Inserting the inflaton's solution in \eqref{phi_Solution}, the second slow-roll parameter $\eta$ is therefore calculated to be
\be
 \eta \equiv \frac{\dot{\epsilon}}{\epsilon H} \simeq - \frac{4 \phi_{,N}}{\phi} 
 = \frac{4 \sqrt{2 \xi}}{H_0 (1 - \epsilon) \sqrt{\lambda} } \phi ~.
\ee
Accordingly, $\phi_\text{end}$ is given by the condition $|\eta| = 1$, i.e.,
\be \label{phi_end}
 \phi_\text{end} \simeq \frac{H_0 (1 - \epsilon) \sqrt{\lambda}}{4 \sqrt{2 \xi}} ~.
\ee
Inserting the expression of $\phi_\text{end}$ \eqref{phi_end} into the solution \eqref{phi_Solution} or the constraint on field range from the $e$-folding number \eqref{Field_Range1}, we acquire the initial value for $\phi_i$ as
\be \label{phi_ini}
 \phi_i \lesssim 0.017 \frac{H_0 (1 - \epsilon) \sqrt{\lambda}}{\sqrt{2 \xi}} ~.
\ee
As a result, by introducing the field range $\Delta \phi \equiv \phi_\text{end} - \phi_i$, it is straightforward to derive the bound on this field range $\Delta\phi$ following Eqs. \eqref{phi_end} and \eqref{phi_ini}:
\be \label{field_range}
 0.23 \frac{H_0 (1 - \epsilon) \sqrt{\lambda}}{\sqrt{2 \xi}} \lesssim \Delta\phi \lesssim \frac{H_0 (1 - \epsilon) \sqrt{\lambda}}{4 \sqrt{2 \xi}} ~.
\ee
Taking the values of the parameters: $\lambda = 2 \times 10^9$, $H_0 = 10^{-5} M_p$, $\xi = 0.1$ and $\epsilon = 0.001$, the approximate field values of \eqref{phi_end} and \eqref{phi_ini} are determined to be $\phi_i \lesssim 0.017 M_p$ and $\phi_\text{end} \simeq 0.25 M_p$, and the field range is given by $0.233 M_p \lesssim \Delta\phi \lesssim 0.25 M_p$.

\subsection{Power spectra} \label{sec:powerspectra}

Due to the narrow resonance effect of the SSR mechanism, primordial density perturbations are exponentially amplified near the characteristic scale $k_*$, while the perturbation modes in the nonresonant regime $k \neq k_*$ behave like the Bunch-Davis vacuum, which is consistent with the scale-independent feature of the primordial density perturbations at the large scales. Thus, at the CMB scales, the power spectra for primordial scalar and tensor perturbations in our model are the same as that of the noncanonical inflation \cite{Garriga:1999vw},
\bl \label{Ps}
 P_\zeta = \frac{1}{8 \pi^2} \frac{H^2}{M_p^2} \frac{1}{c_s \epsilon} ~,~ P_t = \frac{2}{\pi^2} \frac{H^2}{M_p^2} ~,
\el
and their spectra indices are given by \cite{Peiris:2007gz, Bean:2007eh, Baumann:2006cd}
\bl
 n_s - 1 & \equiv \frac{d \ln P_\zeta}{d \ln k} = - 2 \epsilon - 2 \eta - \kappa ~, 
 \\ n_t & \equiv \frac{d \ln P_t}{d \ln k} = - 2 \epsilon ~,
\el
respectively. Scalar perturbations freeze when they exit the sound horizon $c_s k = a H$, while tensor perturbations freeze when they exits the Hubble horizon $k = a H$. In DBI inflation, the scalar spectral index is related to the total $e$-folding number as $n_s - 1 \sim 1/ N_\text{end}$ \cite{Chen:2005ad}, which is consistent with observational data \cite{Ade:2015lrj, Akrami:2018odb}. Additionally, for DBI models, the tensor-scalar ratio on CMB scales is given by \cite{Peiris:2007gz, Bean:2007eh, Baumann:2006cd}
\be
r \equiv \frac{P_t}{P_\zeta} = 16 c_s \epsilon ~.
\ee
which also implies the modified consistency relation \cite{Garriga:1999vw, Lidsey:2006ia}
\be \label{Consistency}
r = - 8 c_s n_t ~.
\ee
All the above formalisms reduce to the cases in the standard canonical inflation scenario when $c_s = 1$.

\begin{figure}[h]
\centering
\includegraphics[width=0.85 \linewidth]{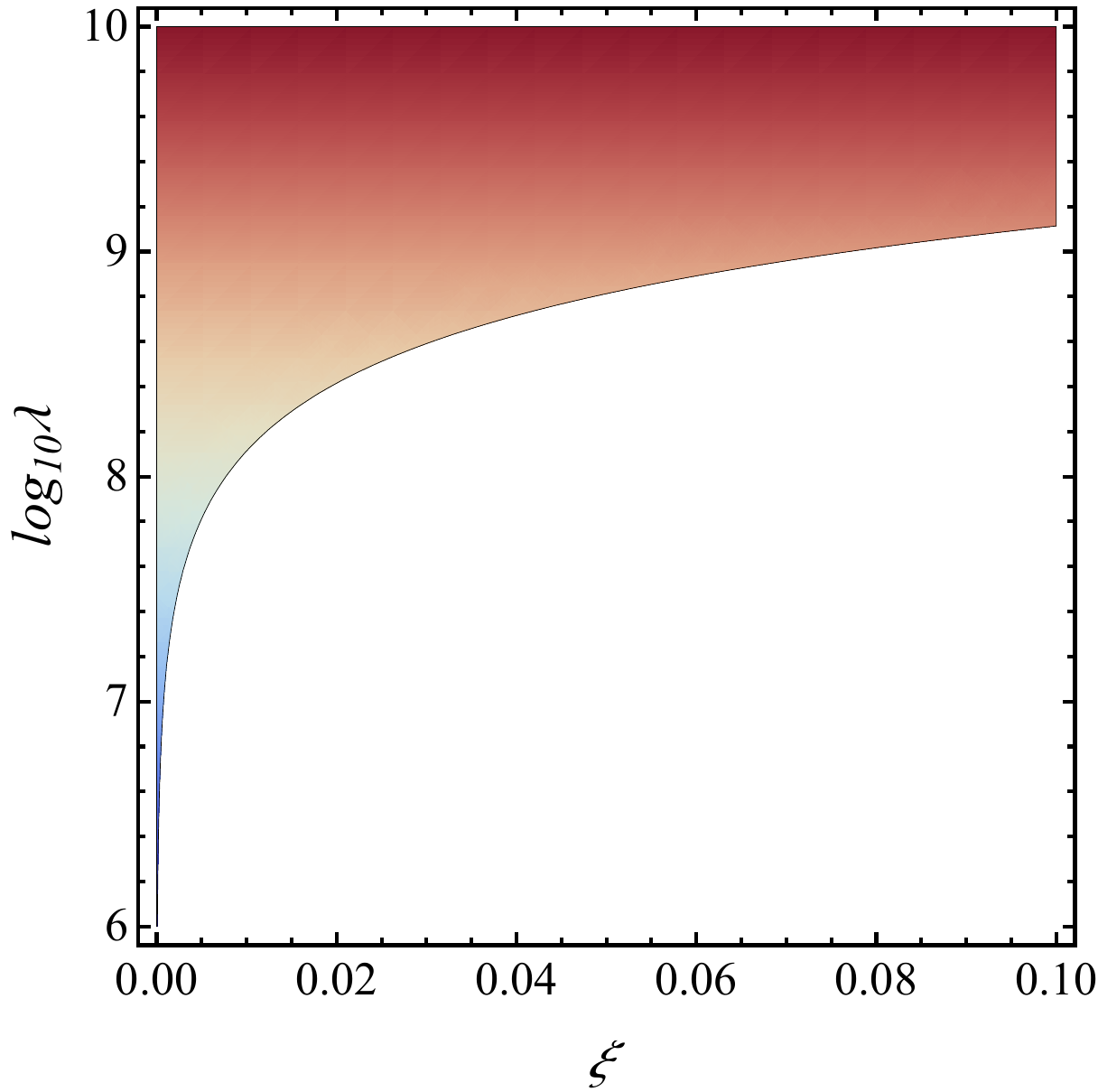}
\caption{The lower bound on $\lambda$ in terms of $\xi$ after adopting the observational fact with $P_\zeta \sim 10^{-9}$. }
\label{fig:lx}
\end{figure}

\begin{figure}[h]
\centering
\includegraphics[width=0.85 \linewidth]{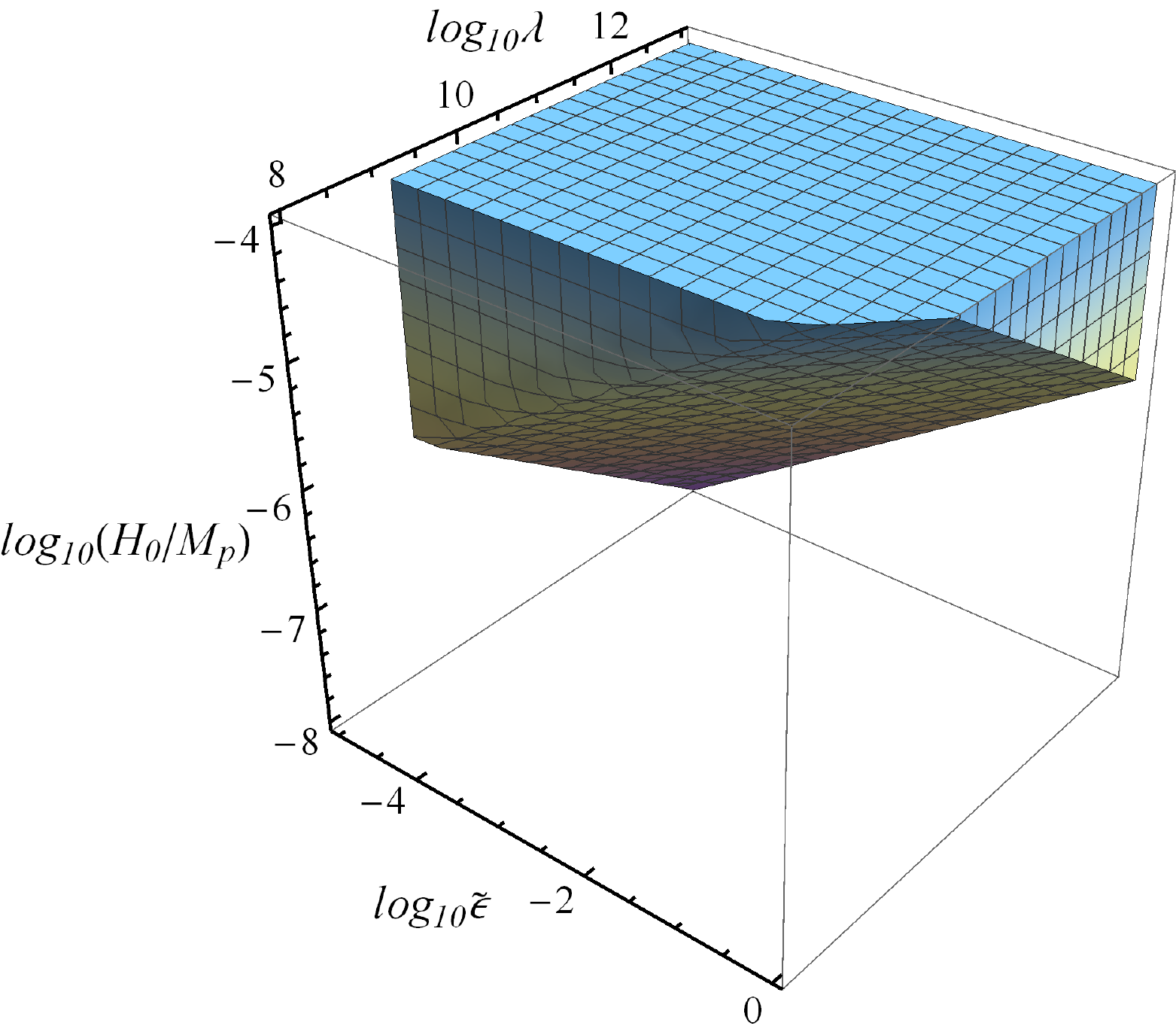}
\caption{The parameter space for $\tilde{\epsilon}$, $H_0$ and $\lambda$ bounded by the constraints in \eqref{slow_bound} and \eqref{Lambda}.}
\label{fig:slow}
\end{figure}

Given the observational fact with $P_\zeta \sim 10^{-9}$, we show below that this can impose a lower bound on the parameter $\lambda$. Since there is $\phi \leq \phi_\text{end}$, we can use Eqs. \eqref{SR} and \eqref{phi_end} and then get,
\be \label{slow_bound}
 \tilde{\epsilon} \lesssim \frac{\lambda}{1024 \xi} \frac{H_0^2}{M_p^2} ~.
\ee
Then, to combine $P_\zeta \sim 10^{-9}$ in Eqs. \eqref{Ps} and \eqref{slow_bound}, we obtain
\be \label{Lambda}
 \lambda \gtrsim 1.3 ~ \xi \times 10^{10} ~.
\ee

Figure \ref{fig:lx} shows the bounds on $\lambda$ in terms of the amplitude $\xi$. For instance, setting $\xi = 0.1$, one gets $\lambda \gtrsim 1.3 \times 10^9$. The relation \eqref{slow_bound} is presented in Fig.~\ref{fig:slow}, providing the bounds on $H_0$ and $\lambda$ in order for $\tilde{\epsilon} < 1$ during inflation. Moreover, if one takes $H_0 = 10^{-5} M_p$ and $\lambda = 2 \times 10^9$, the slow-roll parameter in \eqref{SR} is approximately given by $\tilde{\epsilon} \simeq 0.001$ for primordial power spectra in \eqref{Ps}.

Furthermore, it is known that the Lyth bound of DBI inflation is same as the case in the standard slow-roll inflation \cite{Lyth:1996im, Baumann:2006cd}
\be \label{Lyth1}
\frac{\Delta \phi}{M_p} = \int_{0}^{N_\text{end}} \sqrt{\frac{r}{8}} dN ~.
\ee
Due to Eq.~\eqref{Consistency}, $r$ is a slowly varying small quantity during inflation, the Lyth bound \eqref{Lyth1} is expressed approximately as
\be \label{Lyth2}
\frac{\Delta \phi}{M_p} \simeq \sqrt{\frac{r}{8}} N_\text{end} ~.
\ee
In light of the constraint on the field range \eqref{field_range} and the sufficient $e$-folding number $N_\text{CMB} = 55$, we can find
\be \label{r}
r < 1.5 \times 10^{-4} ~.
\ee
which implies that relic gravitational waves are generally extremely small in our model. We mention that the constraint on $r$ from {\it Planck 2018} data is $r < 0.1$.

\subsection{Primordial non-Gaussianity} \label{sec:NG}

A distinctive theoretical prediction of DBI inflation is the possible large level of primordial non-Gaussianity, with the nonlinear parameter of the equilateral type $f_{NL} \propto c_s^{-2}$ when $c_s \ll 1$. The explicit form of the nonlinear parameter $f_{NL}$ in DBI model \eqref{Action_DBI} is given by \cite{Chen:2006nt}
\be \label{fNL}
f_{NL} = \frac{35}{108} \Big( \frac{1}{c_s^2} - 1 \Big) ~,
\ee
and according to the {\it Planck 2018} experiment \cite{Akrami:2019izv}, the related observational constraint takes: $f_{NL} = -26 \pm 47$ ($68\%$ confidence level), which directly imposes the lower bound on the sound speed squared as follows,
\be \label{cs_Constriant}
c_s^2 \geq 0.015 ~.
\ee
Thus, the phenomenological sound speed squared shown in Fig.~\ref{fig:cs} safely lives within this limit. 

\begin{figure}[h]
\centering
\includegraphics[width=0.9 \linewidth]{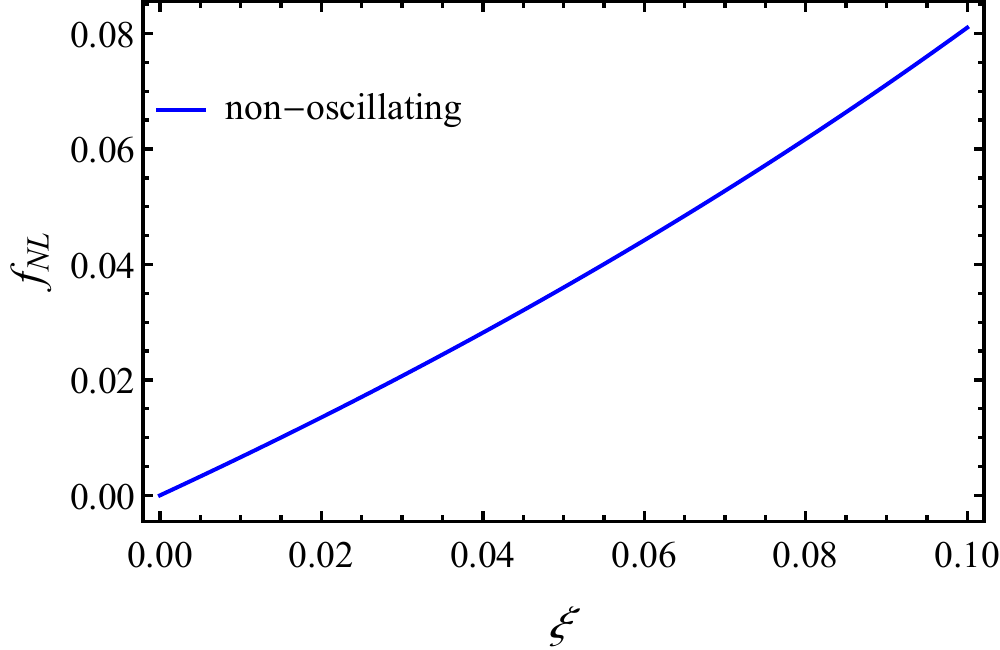}
\caption{The nonlinear parameter $f_{NL}$ as a function of $\xi$ in the nonoscillating stage where $c_s^2 = 1 - 2 \xi$.
}
\label{fig:NG}
\end{figure}

Moreover, we can see below that, the primordial non-Gaussianities predicted by our constructed model are far less than the current observational bounds. In the nonoscillating stage where $c_s^2 = 1 - 2 \xi$, the non-Gaussianity \eqref{fNL} is estimated as
\be
 f_{NL} = \frac{35}{108} \Big( \frac{1}{1 - 2 \xi} - 1 \Big) ~, ~ \text{(nonoscillating)} ~.
\ee
Namely, for $\xi = 0.1$, there is $f_{NL} \simeq 0.081$. Note that, the blue solid curve in Fig.~\ref{fig:NG} shows the dependence of $f_{NL}$ on $\xi$. 

Furthermore, it deserves mentioning that in the oscillating stage $\tau_s < \tau < \tau_\text{end}$, the specific mode of BD vacuum is amplified due to the narrow resonance effect of SSR, and then the common formula \eqref{fNL} is no longer valid. However, as this topic is beyond the scope of the present work, we would like to leave it to be addressed in the follow-up study.


\begin{figure}[h]
\centering
\includegraphics[width=0.9 \linewidth]{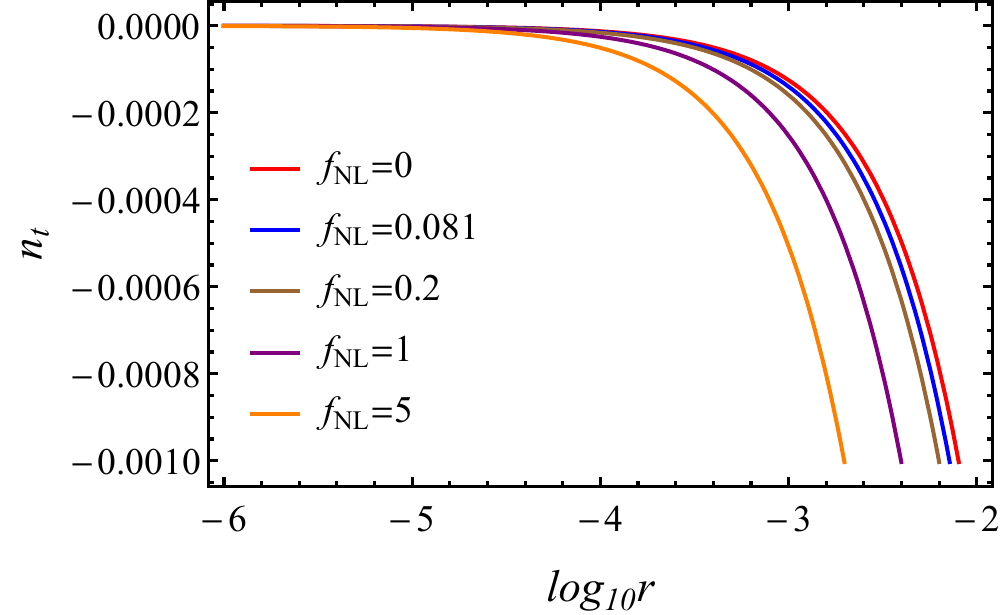}
\caption{The general consistency relation \eqref{r_fNl_nt} with different values of $f_{NL}$ and $\xi$ is fixed to $0.1$. The blue line refers to the corresponding $f_{NL}$ in the nonoscillating stage. The red line represents the standard consistency relation $r = - 8 n_t$.}
\label{fig:r_fNl_nt}
\end{figure}

Using the modified consistency relation \eqref{Consistency} and \eqref{fNL}, we obtain the following generic expression in terms of observables \cite{Lidsey:2006ia, Baumann:2006cd}
\be \label{r_fNl_nt}
 8 n_t = - r \sqrt{1 + \frac{108}{35} f_{NL}} ~.
\ee
Reference \cite{Lidsey:2006ia} has demonstrated that Eq. \eqref{r_fNl_nt} is model independent in the sense that it holds for arbitrary inflaton's potential and warp factor in DBI inflation. Thus, in principle Eq. \eqref{r_fNl_nt} can serve as a smoking gun for DBI inflation with more and more accurate cosmological data. Note that, in the single-field slow-roll inflation, primordial non-Gaussianities are generally quite small \cite{Maldacena:2002vr}, and Eq. \eqref{r_fNl_nt} is exactly the standard consistency relation $r = - 8 n_t$. The general relation \eqref{r_fNl_nt} is shown in Fig.~\ref{fig:r_fNl_nt} in terms of a set of values: $f_{NL} = (0,~ 0.081,~ 0.2,~1,~5)$. One can read that the deviation from the standard consistency relation $r = - 8 n_t$ (i.e. $f_{NL} = 0$ ) becomes larger with the increasing tensor-scalar ratio, while the red tilt of the tensor power spectrum $n_t$ goes up slightly.

\section{Conclusion} \label{sec:conclusion}

In the present study, we have developed the model realization of the SSR mechanism for primordial black hole formation by reconstructing the appropriate form of DBI inflation. In DBI inflation, the noncanonical kinetic term naturally leads to the nontrivial sound speed. Inspired by this feature, we acquire the matching condition for the phenomenological oscillating sound speed in SSR, which is related to a deformed warp factor and the detailed time evolution of inflaton. In order to solve the complicated EoM for DBI inflaton, we have developed a perturbative approach to analyze the background dynamics. The whole inflationary stage is separated into the nonoscillating and the oscillating stages in terms of the evolution of sound speed squared. In the first stage, the sound speed squared is assumed to be a constant slightly deviated from unity, and we have solved the evolution of inflaton by adopting an AdS type of warp factor. Naturally, the inclusion of the oscillating terms in sound speed squared requires a delicately deformation on the AdS-like warp factor. To obtain the form of this warp factor, we restrict the evolution of inflaton to remain almost unchanged, and then the warp factor is solved from the matching condition for the oscillating sound speed. Resorting to the Hamilton-Jacobi formalism, the Hubble parameter and the inflaton's potential are derived. A numerical method is performed to solve the evolution of inflaton, which matches very well with semianalytic results. We also investigate the influence of the nonsmooth conjunction of sound speed at the beginning moment of the oscillating stage on SSR phenomenology, and it turns out that SSR is barely affected by this nonsmooth type of conjunction, and consequently, the same PBH mass spectrum as the case in the original SSR is predicted in our DBI inflation. Regarding the adiabatic condition of the Muhanov-Sasaki equation, we show that the real physical quantity--the comoving curvature perturbation behaves as the BD vacuum in the nonresonant region after Hubble crossing, although the violation of adiabaticity of the Muhanov-Sasaki equation apparently exists.

In light of the {\it Planck 2018} experiment, we derive the constraints on the field range during inflation and show that there exists a quite comparable parameter space of the model to yield the sufficient number of $e$-folds for a successful inflationary phase. By setting the amplitude of primordial density perturbations to be in order of the observed one, the model parameters $\lambda$ can be limited from below and our model typically predicts that the amplitude of primordial gravitational waves is too small to have observable interest. Additionally, the primordial non-Gaussianity predicted in our model depends on the oscillation amplitude of sound speed squared $\xi$, and can easily satisfy the current observational bound. Last but not least, the consistency relation for single-field slow-roll inflation is softly violated in our case due to the small variations of sound speed squared.

\section*{ACKNOWLEDGMENTS}

We are grateful to Jinn-Ouk Gong, Shi Pi, Misao Sasaki, Takahiro Terada, Xi Tong, Dong-Gang Wang, Yi Wang and Masahide Yamaguchi, Sheng-Feng Yan, Pierre Zhang for stimulating discussions. This work is supported in part by the NSFC (No. 11722327, No. 11961131007, No. 11653002 and No. 11421303), by the CAST Young Elite Scientists Sponsorship (2016QNRC001), by the National Youth Talents Program of China, and by the Fundamental Research Funds for Central Universities. All numerics are operated on the computer clusters {\it LINDA \& JUDY} in the particle cosmology group at USTC. C.C. is grateful to Weixia Chen\&Xueying Tian for their hospitality and support.

\appendix

\section{The Precise Solution for Inflaton $\phi(\tau)$}

\label{app:check}

We take the approximation that $H$ is treated as a constant when we solve the matching condition \eqref{Matching_2}, and obtain the approximated solution \eqref{phi_tau}. In this Appendix, we will solve $\phi(\tau)$ rigorously and compare the precise solution with the approximated one. 

\begin{figure}[h]
	\centering
	\includegraphics[width=3in]{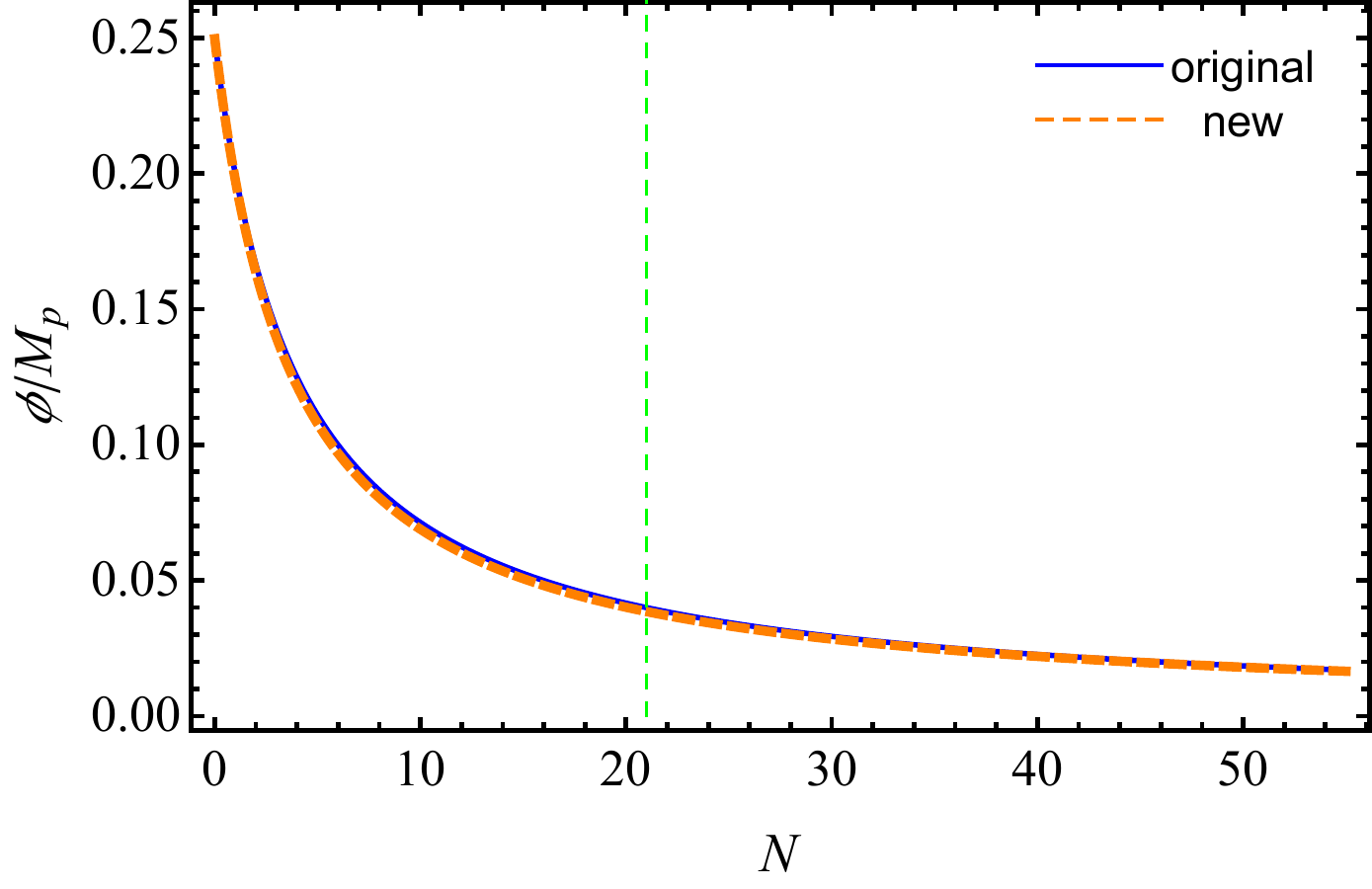}
	\caption{The approximated solution \eqref{phiN} (the blue curve) and the precise solution \eqref{phi_N} (the orange dashed curve) of the evolutions of inflaton field $\phi$. The green dashed line denotes the beginning of the oscillating stage. The parameter values are chosen to be: $\lambda = 2 \times 10^{9}$, $H_0 = 10^{-5} M_p$, $\xi = 0.1$ and $N_s = 21$.}
	\label{fig:1}
\end{figure}

Starting from the definition of the slow-roll parameter $\epsilon \equiv - \dot{H} / H^2$ and the assumption that $\epsilon$ is regarded as a constant in the quasi-de Sitter expansion, we can yield three equivalent expressions for scale factor in the conformal time
\be \label{atau1}
a(\tau) = a_0^{ { \epsilon \over \epsilon -1 } } { 1 \over [- \tau H_0 (1 - \epsilon)]^{1 \over 1 - \epsilon} } ~,
\ee
and
\be \label{atau2}
a(\tau) = \frac{1}{\tau H(\tau) (\epsilon - 1)} ~,
\ee
and
\be \label{atau3}
a(\tau) = a_0 \l( { \tau \over \tau_i }\r)^{ 1 \over \epsilon -1 } ~,
\ee
Note that the second one \eqref{atau2} is what we used in this paper. The Hubble parameter $H_0$ and the scale factor $a_0$ are valuated at the initial time $\tau_i$ which is set to be the beginning moment of inflation. For the purpose of yielding the precise solution of $\phi(\tau)$, it is convenient to adopt the last expression \eqref{atau3} for the scale factor. Plugging \eqref{atau3} into the matching condition \eqref{Matching_2}, we yield
\be \label{phi_tau2}
\phi(\tau)
=
\Big[ \frac{1}{\phi_i} + \sqrt{ 2\xi \over \lambda } { 1 \over \epsilon H_0} \Big( 1 - { \Big( { \tau \over \tau_i } \Big) }^{ \epsilon \over \epsilon - 1} \Big) \Big]^{-1} ~.
\ee
which can also be written in terms of $e$-folding number
\be \label{phi_N}
\phi(N) = \Big[ \frac{1}{\phi_\text{end}} + \sqrt{ 2\xi \over \lambda } { 1 \over \epsilon H_0} \Big( e^{ \epsilon N_\text{end} } - e^{ \epsilon ( N_\text{end} - N ) } \Big) \Big]^{-1} ~,
\ee
Note that only the increasing solution is remained as for IR DBI model. It is straightforward to check that, the leading order of the precise solution \eqref{phi_N} in terms of $\epsilon$ is same as our original solution \eqref{phi_tau} which can be written as
\bl \label{phiN}
\phi(N)
&\simeq \Big[ \frac{1}{\phi_i} + \frac{\sqrt{2\xi}}{H_0 (1 - \epsilon) \sqrt{\lambda}} (N - N_\text{end}) \Big]^{-1} \nn
\\&
\simeq
\Big[ \frac{1}{\phi_\text{end}} + \frac{\sqrt{2\xi}}{H_0 (1 - \epsilon) \sqrt{\lambda}} N \Big]^{-1} ~.
\el
Figure \ref{fig:1} shows a comparison between these two results, it is very clear that our original solution is reasonably good for the description of inflaton evolution.

\bibliography{ssr_DBI}

\end{document}